\documentclass[10pt,journal,compsoc]{IEEEtran}
\usepackage{graphicx}
\usepackage{caption}
\ifCLASSOPTIONcompsoc
  \usepackage[nocompress]{cite}
\else
  \usepackage{cite}
\fi
\ifCLASSINFOpdf
\else
\fi

\usepackage{amssymb}
\usepackage{balance}
\usepackage{bm}
\usepackage{adjustbox}
\usepackage{graphicx,xcolor} 
\usepackage[flushleft]{threeparttable}
\usepackage{balance}
\usepackage{algorithm}
\usepackage{algpseudocode}
\usepackage{multirow}
\usepackage{booktabs}
\usepackage{listings}
\usepackage{xcolor}
\usepackage{amsmath}
\usepackage{algpseudocode}
\usepackage{tabularx,booktabs}

\newcolumntype{Y}{>{\centering\arraybackslash}X}

\algblock{ParFor}{EndParFor}
\algnewcommand\algorithmicparfor{\textbf{parfor}}
\algnewcommand\algorithmicpardo{\textbf{do}}
\algnewcommand\algorithmicendparfor{\textbf{end\ parfor}}
\algrenewtext{ParFor}[1]{\algorithmicparfor\ #1\ \algorithmicpardo}
\algrenewtext{EndParFor}{\algorithmicendparfor}

\definecolor{codegreen}{rgb}{0,0.6,0}
\definecolor{codegray}{rgb}{0.5,0.5,0.5}
\definecolor{codepurple}{rgb}{0.58,0,0.82}
\definecolor{backcolour}{rgb}{0.98,0.98,0.95}

\lstdefinestyle{mystyle}{
    backgroundcolor=\color{backcolour},   
    commentstyle=\color{codegreen},
    keywordstyle=\color{magenta},
    numberstyle=\tiny\color{codegray},
    stringstyle=\color{codepurple},
    basicstyle=\ttfamily\footnotesize,
    breakatwhitespace=false,         
    breaklines=true,                 
    captionpos=b,                    
    keepspaces=true,                 
    numbers=left,                    
    numbersep=5pt,                  
    showspaces=false,                
    showstringspaces=false,
    showtabs=false,                  
    tabsize=2
}

\begin{document}
%
\title{HitGNN: \underline{Hi}gh-\underline{t}hroughput \underline{GNN} Training Framework on CPU+Multi-FPGA Heterogeneous Platform}
%
%
%
%

\author{Yi-Chien~Lin,
        Bingyi~Zhang,
       Viktor~Prasanna
       
\IEEEcompsocitemizethanks{\IEEEcompsocthanksitem Y. Lin, B. Zhang and V. K. Prasanna are with the Department of Electrical and
Computer Engineering, University of Southern California, Los Angeles,
CA 90089. 
E-mail: \{yichienl, bingyizh, prasanna\}@usc.edu}       
       
}

%
%

\markboth{Journal of \LaTeX\ Class Files,~Vol.~14, No.~8, August~2015}%
{Shell \MakeLowercase{\textit{et al.}}: Bare Demo of IEEEtran.cls for Computer Society Journals}
%



\IEEEtitleabstractindextext{%
\begin{abstract}
As the size of real-world graphs increases, training Graph Neural Networks (GNNs) has become time-consuming and requires acceleration.
While previous works have demonstrated the potential of utilizing FPGA for accelerating GNN training, few works have been carried out to accelerate GNN training with multiple FPGAs due to the necessity of hardware expertise and substantial development effort. 
To this end, we propose HitGNN, a framework that enables users to effortlessly map GNN training workloads onto a CPU-Multi-FPGA platform for acceleration.
In particular, HitGNN takes the user-defined synchronous GNN training algorithm, GNN model, and platform metadata as input, determines the design parameters based on the platform metadata, and performs hardware mapping onto the CPU+Multi-FPGA platform, automatically.
HitGNN consists of the following building blocks: 
(1) high-level application programming interfaces (APIs) that allow users to specify various synchronous GNN training algorithms and GNN models with only a handful of lines of code;
(2) a software generator that generates a host program that performs mini-batch sampling, manages CPU-FPGA communication, and handles workload balancing among the FPGAs;
(3) an accelerator generator that generates GNN kernels with optimized datapath and memory organization.
We show that existing synchronous GNN training algorithms such as DistDGL and PaGraph can be easily deployed on a CPU+Multi-FPGA platform using our framework, while achieving high training throughput.
Compared with the state-of-the-art frameworks that accelerate synchronous GNN training on a multi-GPU platform, HitGNN achieves up to 27.21$\times$ bandwidth efficiency, and up to 4.26$\times$ speedup using much less compute power and memory bandwidth than GPUs.
In addition, HitGNN demonstrates good scalability to 16 FPGAs on a CPU+Multi-FPGA platform.
\end{abstract}

}

\maketitle

\IEEEdisplaynontitleabstractindextext

%
\IEEEpeerreviewmaketitle

\IEEEraisesectionheading{\section{Introduction}\label{sec:introduction}}
\IEEEPARstart{G}{raph} Neural Networks (GNNs) have become the state-of-the-art models for representation learning on graphs, facilitating many applications such as social recommendation system \cite{recommend1,recommend2}, molecular property prediction \cite{graphsage,yang_li_2023} and traffic prediction \cite{traffic}, etc. Initially, GNNs were computed on a GPU \cite{graphsaint, shaDow, clustergcn} or an FPGA platform \cite{hp-gnn,lukGCN,linGCN}; however, as the size of the graph increases, computing GNNs on a single GPU or an FPGA platform becomes time-consuming. 
Thus, many works \cite{p3, distdgl, 2p, pagraph, gnnlab} have proposed to accelerate GNN training on a multi-CPU or a multi-GPU platform as it provides more memory bandwidth and computation resources. 
To train GNN on multiple devices in parallel, these works perform \textit{synchronous GNN training}; we describe synchronous GNN training in detail in Section \ref{sec:method}.

Compared with general-purpose processors like CPU and GPU, CPU+FPGA heterogeneous platform is promising for GNN training acceleration:
the CPU can flexibly support various graph preprocessing and mini-batch sampling algorithms; and the FPGA can efficiently perform GNN operations because FPGA supports customized data access pattern and memory organization, which can effectively reduce the substantial memory traffic and random memory accesses in GNN training. 
Despite the various optimizations that can be deployed on a CPU+FPGA platform, training GNNs on a CPU+FPGA platform can still be time-consuming due to limited computation power and memory resources; thus, it is desirable to accelerate GNN training on a CPU+Multi-FPGA heterogeneous platform.
However, accelerating GNN training on a CPU+Multi-FPGA platform is challenging. 
First, training GNNs on a CPU+Multi-FPGA platform suffers from workload imbalance and significant data communication overhead among FPGAs, which leads to low performance and poor scalability.
Second, it requires hardware expertise to develop optimized kernels and considerable amount of time to explore the complex hardware design space of a CPU+Multi-FPGA platform. 

Motivated by these challenges, we propose HitGNN, a generic framework for mapping synchronous GNN training algorithms on a CPU+Multi-FPGA heterogeneous platform.
We first formulate the high-level abstraction of synchronous GNN training algorithms and then develop HitGNN based on the abstraction; this allows our framework to support various training algorithms that can be described with the formulated abstraction. 
To achieve high throughput and automate the implementation process, we develop a hardware Design Space Exploration (DSE) engine. 
\begin{figure}[h]
    \centering
    \includegraphics[width=8cm]{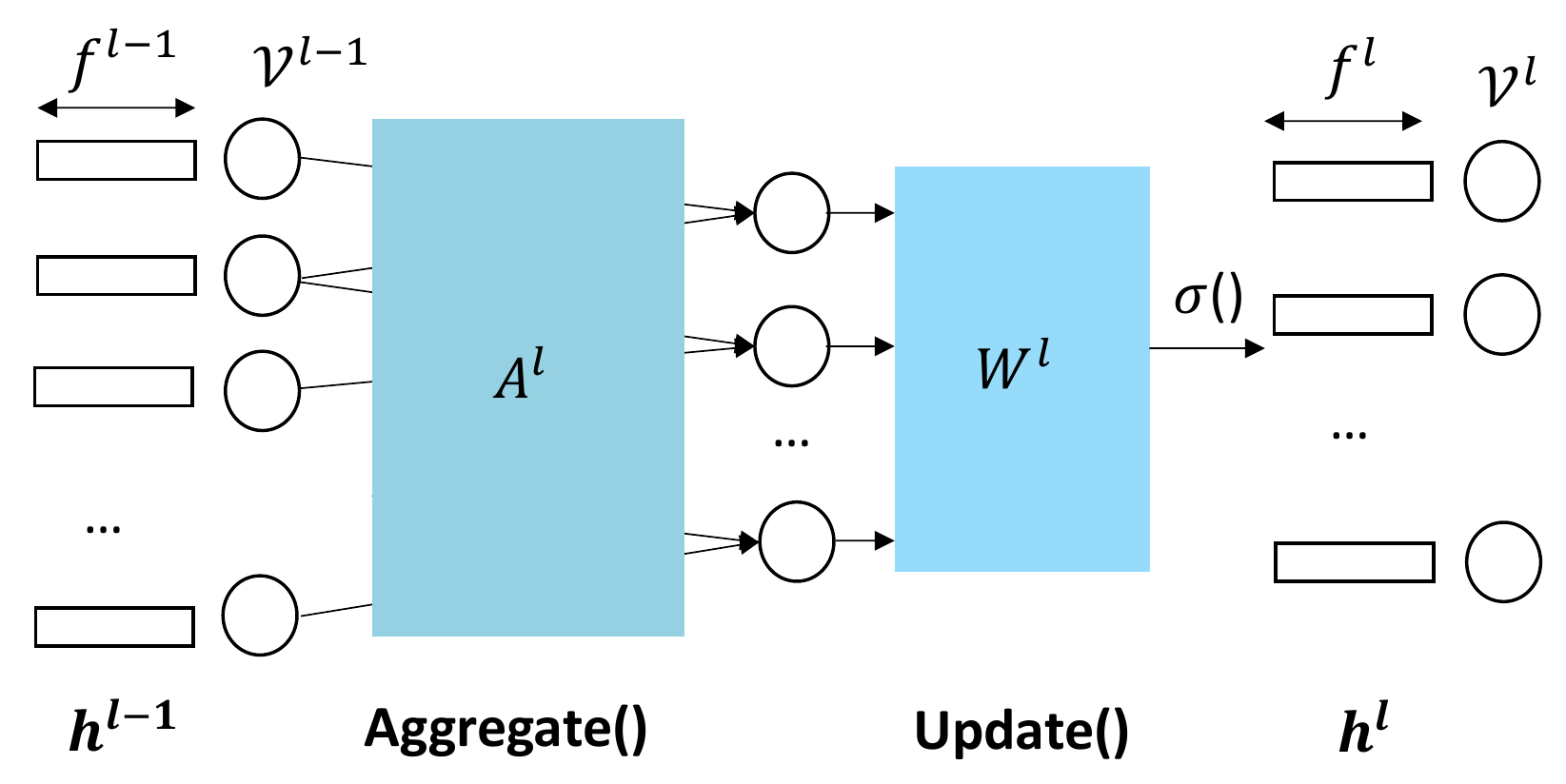}
    \caption{Computation abstraction of a GNN layer}
     \label{fig:gnn}
\end{figure} 
Given the platform metadata, the DSE engine determines the accelerator configurations that optimize the training throughput.
To reduce the development effort, HitGNN features a kernel library, which consists of optimized GNN kernels that can be used off-the-shelf.
To mask the hardware implementation details, HitGNN provides application programming interfaces (APIs) that allow developers to easily implement various training algorithms and GNN models with only a handful of lines of code.
Lastly, the host program of HitGNN performs mini-batch sampling, handles workload imbalance, and reduces data communication among FPGAs.

We summarize our contributions as follows:
\begin{itemize}
    \item We propose HitGNN, a generic framework that can automatically map various synchronous GNN training algorithms such as P$^3$, DistDGL and PaGraph, and various GNN models such as GCN and GraphSAGE on a CPU+Multi-FPGA heterogeneous platform for acceleration. 
    \item To abstract away the hardware implementation details, HitGNN features easy-to-use programming APIs, allowing users to specify various synchronous GNN training algorithms and GNN models with a handful of lines of software code.
    \item To reduce the development effort, we design a GNN kernel library that consists of parameterized and optimized kernels of various well-known GNN models, and a DSE engine that can automatically determine the accelerator configuration.
    \item {To realize a scalable design,}  we develop optimizations to reduce FPGA-to-FPGA communication overhead, and balance the workload among the FPGAs.
    \item Compared with state-of-the-art synchronous GNN training implementations on a multi-GPU platform, HitGNN achieves up to 27.21$\times$ bandwidth efficiency, and 4.26$\times$ speedup using much less computation resources and memory bandwidth.
\end{itemize}

\section{Background}
\subsection{GNN Models} \label{GNN_model}
Given an input graph $\mathcal{G}(\mathcal{V}, \mathcal{E}, \bm{X})$, where $\mathcal{V}$, $\mathcal{E}$, and $\bm{X}$ denote the set of vertices, the set of edges, and the feature matrix of the graph, respectively,
a GNN model is specified by:
\begin{itemize}
    \item $L$: number of layers.
    \item $\mathcal{V}^{t}$: a set of target vertices to be inferred.
    \item $f^{l}$: hidden dimension of layer $l~( 1 \leqslant l \leqslant L)$.
    \item A mechanism to construct mini-batches, including: 
\begin{algorithm}
\caption{GNN Computation Abstraction}
\label{alg:aggregate-update-paradigm-define}

\begin{algorithmic}[1]
\For{$l=1...L$}
\For{vertex $v \in \mathcal{V}^l$}
\State{$\bm{a}^l_{v} = \textbf{Aggregate(}\bm{h}_{u}^{l-1}: u\in \mathcal{N}(v)$ and $u\in \mathcal{V}^{l-1}\textbf{)}$}
\State{$\bm{h}_{v}^l = \textbf{Update(}\bm{a}_i^l, \bm{W}^{l}, \sigma() \textbf{)}$}
\EndFor
\EndFor
\end{algorithmic}
\end{algorithm}
    \begin{itemize}
        \item The mechanism to construct $\mathcal{V}^{l}$: the set of vertices in layer $l~( 0 \leqslant l \leqslant L) $. $|\mathcal{V}^{l}|$ denotes the number of vertices in layer $l$. Moreover, $\mathcal{V}^{L} = \mathcal{V}^{t}$.
         \item The mechanism to construct $\bm{A}^{l} \in \mathbb{R}^{{|\mathcal{V}^{l-1}|}\times {|\mathcal{V}^{l}|}} $: adjacency matrix for feature aggregation in layer $l~( 1 \leqslant l \leqslant L)$. $\bm{A}^{l}$ defines the inter-layer connectivity (edges) between $\mathcal{V}^{l-1}$ and $\mathcal{V}^{l}$. 
    \end{itemize}
   \item \textbf{Aggregate(\hspace{0.05cm})} function that is used by each vertex to aggregate information from its neighbors.
    \item \textbf{Update(\hspace{0.05cm})} function including a multi-layer perceptron (MLP) and an activation function $\sigma$(\hspace{0.05cm}) that is used to perform feature update.
    \item $\bm{W}^{l}\in \mathbb{R}^{f^{l-1}\times f^{l}}$: weight matrix of layer $l~( 1 \leqslant l \leqslant L)$ that is used in update function to perform linear transformation of vertex features.
    \item $\bm{X}\in \mathbb{R}^{{|\mathcal{V}|}\times f^{l}}$: input feature matrix where each row represents the feature vector of a vertex.
    \item $\bm{h}^l \in \mathbb{R}^{{|\mathcal{V}^{l}|}\times f^{l}}$: the vertex feature matrix in layer $l~( 0 \leqslant l \leqslant L)$. Moreover, the feature matrix of the input layer is the input feature matrix, i.e., $\bm{h}^0 = \bm{X}$; and the feature matrix of the last layer $\bm{h}^L$ is the node embeddings of the target vertices $\mathcal{V}^t$.
\end{itemize}
GNN learns to generate low-dimensional vector representation (i.e., node embedding) for a set of target vertices $\mathcal{V}^{t}$ by iteratively aggregating and updating the vertex features from their $L-$hop neighbors. 
We depict the computation abstraction of a GNN layer in Figure \ref{fig:gnn}. 
Starting from layer $1$, the GNN model computes the feature vector of each vertex in $\mathcal{V}^{1}$ by aggregating and updating the feature vectors of its neighbor vertices in $\mathcal{V}^{0}$; this process is repeated $L$ times until the node embeddings of the target vertices $\mathcal{V}^{t}$ (which is $\mathcal{V}^{L}$) are derived.
The computation process of a GNN model is shown in Algorithm \ref{alg:aggregate-update-paradigm-define}, which is also known as the aggregate-update paradigm \cite{hygcn}. $\bm{a}_{v}^l \in \mathbb{R}^{f^{l}} $ is the intermediate result of $v\in \mathcal{V}^{l}$, and $\mathcal{N}_{s}(v)$ denotes sampled neighbors of $v$ in $\mathcal{V}^{l-1}$.

\subsection{Mini-batch GNN Training}
GNN models can be trained in mini-batch fashion, the training process consists of five stages \cite{graphsage, graphsaint, clustergcn}: sampling, forward propagation, loss calculation, back propagation and weight update. In the sampling stage, a set of vertices and adjacency matrices are sampled from the input graph topology $\mathcal{G}(\mathcal{V}, \mathcal{E})$.
We use $\mathcal{V}^l$ to denote the vertices sampled from $\mathcal{V}$ in layer $l$. $\bm{A}^l$ denotes the sampled adjacency matrix, which describes inter-layer connections (edges) between $\mathcal{V}^{l-1}$ and $\mathcal{V}^{l}$ within the mini-batch. A mini-batch consists of target vertices $\mathcal{V}^L$, sampled vertices for each layer $\{\mathcal{V}^{l}:0\leqslant l\leqslant L-1\}$, and sampled adjacency matrices (edges) $\{\bm{A}^l:1\leqslant l\leqslant L-1\}$. 

\begin{table*}[]
\centering
\caption{Synchronous GNN training algorithms}
\label{tab:algo}
\begin{tabular}{l|l|l}
\toprule
\begin{tabular}[c]{@{}l@{}}Synchronous GNN \\ Training Algorithm\end{tabular} & Graph Partitioning                                                                                               & Feature Storing Strategy                               \\ \midrule \midrule
DistDGL \cite{distdgl}                                                                       & METIS with multi-constraints                                                                                     & Based on the graph partitioning                        \\ \midrule
PaGraph \cite{pagraph}                                                                       & \begin{tabular}[c]{@{}l@{}}A greedy approach which aims to  balance \\ the number of training vertices among partitions\end{tabular} & Store feature vectors of vertices with high out-degree \\ \midrule
P$^3$ \cite{p3}                                                                             & Partition along the feature dimension                                                                            & Based on the graph partitioning                        \\ \bottomrule
\end{tabular}
\end{table*}

\begin{small}
\begin{algorithm}
\caption{Mini-batch GNN Training Algorithm}
\label{alg:GNN}
\begin{algorithmic}[1]
\For{each iteration}
\State{\textbf{Sampling($\mathcal{G}(\mathcal{V}, \mathcal{E})$)}} 
{\color{blue}\Comment{Derive mini-batches}}
\For{$l=1...L$}
{\color{blue}\Comment{Forward Propagation}}
\For{vertex $v \in \mathcal{V}^l$}
\State{$\bm{a}^l_{v} = \textbf{Aggregate(}\bm{h}_{u}^{l-1}: u\in \mathcal{N}_{s}(v)$, $u\in \mathcal{V}^{l-1}\textbf{)}$}
\State{$\bm{h}_{v}^l = \textbf{Update(}\bm{a}_i^l, \bm{W}^{l}, \sigma() \textbf{)}$}
\EndFor
\EndFor
\State{\textbf{CalculateLoss($\{\bm{h}_{i}^{L}:v_{i}\in \mathcal{V}^{L}\}$)}}
\State{\textbf{BackPropagation( )}}
{\color{blue}\Comment{Derive gradient of ${W}^l$}}
\State{\textbf{WeightUpdate( )}}
\EndFor
\end{algorithmic}
\end{algorithm}
\end{small}
In the forward propagation stage, the mini-batch is processed layer by layer; the output of the last layer is the node embeddings of the target vertices $\{\bm{h}_{v}^{L}:v\in \mathcal{V}^{L}\}$, which are then compared with the ground truth for loss calculation. The calculated loss served as the input for back propagation, which performs a similar computation as forward propagation but in the reverse direction. Finally, the gradients of $\bm{W}^l$ in each layer are derived for weight update. We show the steps of GNN training in Algorithm \ref{alg:GNN}. 

GNN models can also be trained using full-graph, this approach does not require the sampling stage; 
however, full-graph training causes large memory footprint \cite{FaaS, understand_GNN} that may not fit in a device memory (e.g., FPGA local DDR).
Therefore, HitGNN focuses on accelerating mini-batch GNN training as it demonstrates advantages in accuracy, scalability on large graphs, and has been adopted by many state-of-the-art GNN frameworks \cite{graphsaint, pyg, pagraph, clustergcn}.

\subsection{Synchronous GNN Training}\label{sec:method}

Existing works \cite{distdgl, p3, pagraph} utilize Synchronous Stochastic Gradient Descent (SGD) \cite{sgd} to train GNN on a multi-CPU or multi-GPU platform. 
For the rest of the paper, we use \textit{synchronous GNN training} to refer to training GNN on multiple devices in parallel using synchronous SGD. 

Synchronous GNN training is similar to Algorithm \ref{alg:GNN}, but with two additional stages: graph preprocessing and gradient synchronization.
The first stage is graph preprocessing. In this stage, the input graph $\mathcal{G}(\mathcal{V}, \mathcal{E})$ is partitioned and distributed to each device such as GPU or FPGA for parallel training.
In addition to graph partitioning, the graph preprocessing stage also performs 
feature storing which stores feature vectors in the device local memory (e.g., GPU global memory or FPGA local DDR).
For devices like GPU or FPGA, the entire feature matrix $\bm{X}$ of a large-scale graph may be too large to fit in the local memory; 
thus, existing works \cite{distdgl,p3,pagraph} develop various feature storing strategies to store only part of the feature matrix in the device local memory.
\begin{figure}[h]
    \centering
    \includegraphics[width=8.5cm]{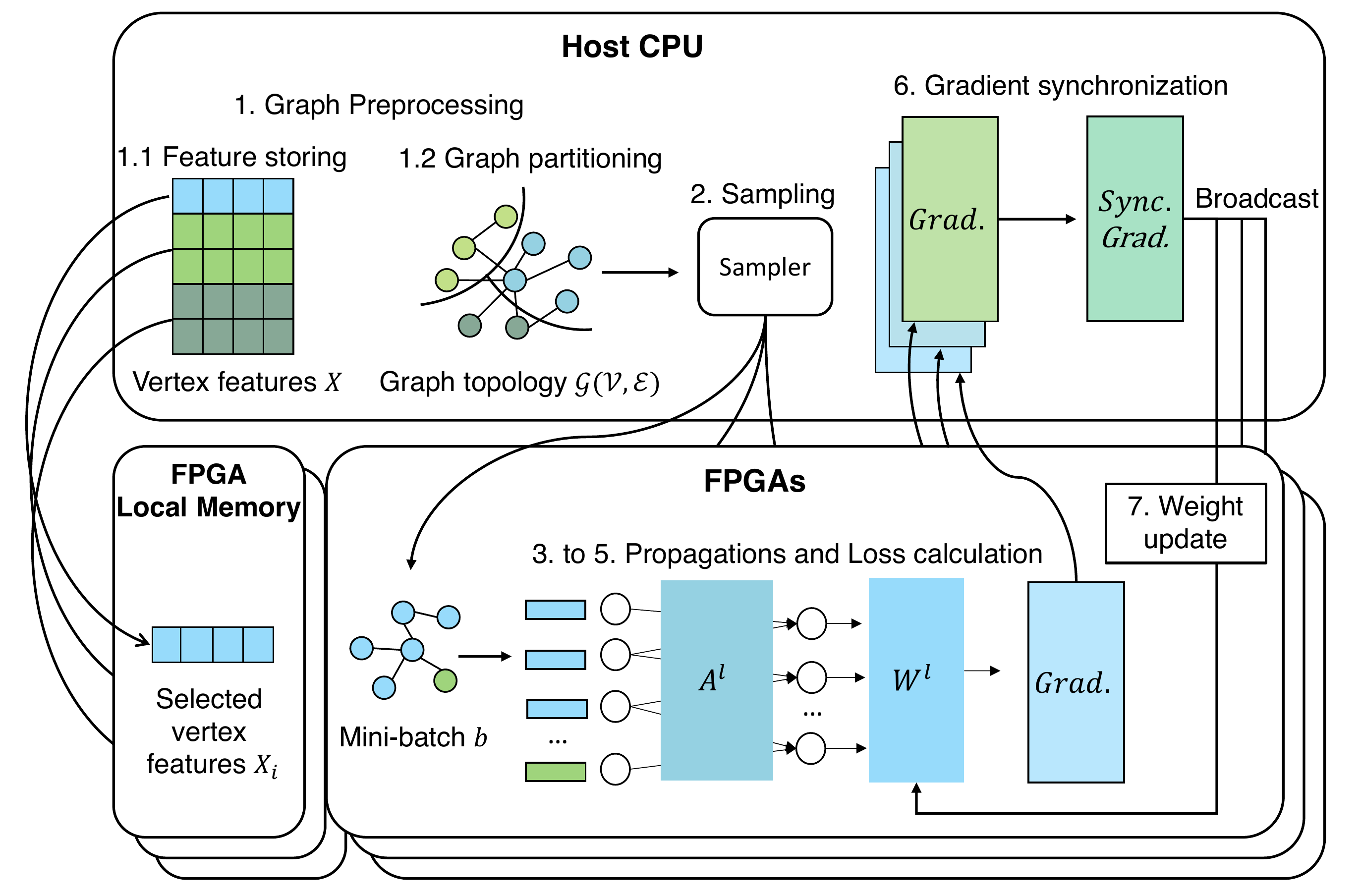}
    \caption{Synchronous GNN training on a CPU+Multi-FPGA platform}
    \vspace{-0.3cm}
     \label{fig:method}
\end{figure}
We use $\bm{X}_i$ to denote the selected feature vectors stored in the local memory of device $i$.
For works like DistDGL, the feature storing strategy is based on the result of graph partitioning; i.e., if $v_j$ belongs to partition $i$, then the feature vector of $v_j \in X_i$. 
Other works like PaGraph develop caching strategies to store feature vectors of frequently accessed vertices, which is independent of the graph partitioning. 
When the graph preprocessing is done, each device performs forward propagation, loss calculation, and back propagation in parallel. 
Then, a gradient synchronization is performed, which averages the gradients collected from each device. 
Then, the averaged gradient is broadcast to update the model weight within each device.
We depict the workflow of synchronous GNN training on a CPU+Multi-FPGA platform in Figure \ref{fig:method}.
{The CPU+Multi-FPGA platform consists of a single CPU or multiple CPUs (depending on the number of sockets in the machine), and multiple FPGAs.
The CPUs are connected to the FPGAs via PCIe
and each FPGA is connected to a local DDR memory.}


We list several representative synchronous GNN training algorithms in Table \ref{tab:algo}. The differences among these algorithms are in graph partitioning and feature storing strategy. Other stages such as forward propagation, gradient synchronization, etc. are identical. Thus, we only show the two different stages for simplicity.

\begin{figure*}[h]
    \centering
    \includegraphics[width=15cm]{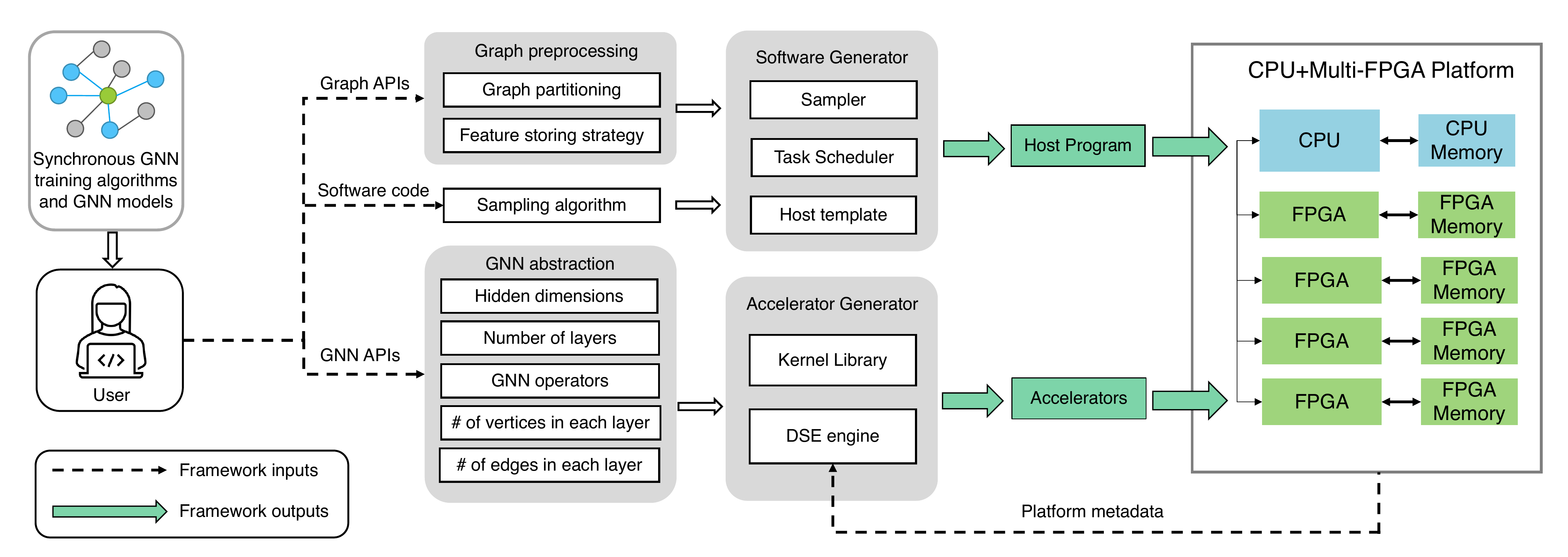}
    \caption{Framework overview}
     \label{fig:framework}
\end{figure*} 

\subsection{Related Work}
{Several works have been proposed to accelerate neural network training using multiple FPGAs: 
\cite{CNN2, geng2018framework} accelerate Convolutional Neural Network (CNN) training using an FPGA cluster.
While these works show promising results in terms of performance and energy-efficiency, CNN accelerators cannot be directly adapted to GNN training. 
This is because the computation characteristics of CNN and GNN are quite different: CNN models feature structured input data with high computation intensity, while GNN models feature unstructured input data with low computation intensity.
There are also works that accelerate GNN training using multiple FPGAs.
\cite{FaaS} accelerates GNN training on a distributed platform, where the graph is stored in multiple nodes. 
On a distributed platform, the training performance is bottlenecked by the sampling stage.
In this work, we focus on a single-node platform with multiple attached FPGAs.
On such a platform, the performance is bottlenecked by the propagation stage. 
\cite{CARLA} accelerates GNN training on a single-node platform;
however, \cite{CARLA} only accelerates a specific synchronous GNN training algorithm and does not provide any tool for users to reduce development effort.
In this work, we propose a general framework that can accelerate various synchronous GNN training algorithms on a CPU+Multi-FPGA platform. 
Furthermore, our framework features software APIs, a DSE engine, and an optimized library to reduce the development effort of accelerating GNN training on a CPU+Multi-FPGA platform.}

\section{{Challenges}}\label{sec:challenge}

We identify several challenges in accelerating synchronous GNN training on a CPU+Multi-FPGA platform.

\vspace{0.1cm}
\noindent\textbf{Challenge 1:} Developing a GNN training accelerator on a CPU+Multi-FPGA platform is time-consuming, and requires hardware expertise.
In particular, the user needs to develop highly-optimized kernels and explore the hardware design space to fully utilize the available resources; the user also needs to develop a host program to manage task scheduling and data communication among FPGAs.

\vspace{0.1cm}
\noindent\textbf{Challenge 2:} 
{Achieving scalable speedup on a CPU+Multi-FPGA platform.}
First, the data dependency of the graph-structured data incurs high FPGA-to-FPGA communication overhead during training.
Second, the workload on each FPGA may be imbalanced due to graph partitioning.
For example, in DistDGL \cite{distdgl}, METIS aims to minimize the cross-partition edge connections to reduce communication overhead, but it cannot assign the same number of vertices and edges to each partition and thus leads to workload imbalance.

\vspace{0.1cm}
\noindent\textbf{Challenge 3:} The optimizations in HitGNN need to be general enough to support various synchronous GNN training algorithms; also, the optimizations should not alter the algorithm itself as it would affect the model accuracy and convergence rate.

To address Challenge 1, HitGNN abstracts away the hardware implementation details via software-programming APIs, and automates the design steps via the Software Generator and the Accelerator Generator (Section \ref{sec:overview}); 
to address Challenges 2 and 3, we develop several optimizations for the CPU+Multi-FPGA platform to increase scalability; these optimizations can be applied to various synchronous GNN training algorithms. 
In addition, even though we applied several optimizations, HitGNN performs the same computations as in the original training algorithm; thus, the accuracy and convergence rate remain the same (Section \ref{sec:opt}).

\section{Framework}\label{sec:overview}

We describe the workflow of HitGNN in the following:
in the design phase, user specifies the synchronous GNN training algorithm, GNN model, and the target platform metadata. 
Then, HitGNN automatically generates optimized accelerator designs and a host program (Section \ref{sec:framework}); during the runtime phase, user launches the GNN training on the CPU+Multi-FPGA platform. We show how the generated designs are mapped to the target platform in Section \ref{sec:system}.

\subsection{{Framework Overview}}\label{sec:framework}
We depict the framework overview of HitGNN in Figure \ref{fig:framework}.
HitGNN takes a synchronous GNN training algorithm, a GNN model, and platform metadata as input, and generates a high-throughput design to accelerate the training on a CPU+Multi-FPGA platform. 
In particular, the design consists of (1) a host program that manages task scheduling, data communication, and mini-batch sampling; and (2) accelerator designs which are optimized GNN kernels that run on the FPGA.
In the input program, user specifies a synchronous GNN training algorithm via two sets of APIs:
\begin{itemize}
    \item \textbf{Graph APIs}: specify the graph partitioning and feature storing strategy for the graph preprocessing stage mentioned in Section \ref{sec:method}.
    \item \textbf{GNN APIs}: parameters that define a GNN model mentioned in Section \ref{GNN_model}. 
\end{itemize}
\renewcommand{\arraystretch}{0.85}
\begin{table*}[]
\caption{Application Programming Interfaces of HitGNN}
\label{tab:API}
\centering
\begin{tabular}{@{}l|l|l@{}}
\toprule
API Type                    & API Functions             & Description \\ \midrule
\multirow{2}{*}{Graph APIs} & Graph\_Partition( ) &   Assign a set of vertices $\mathcal{V}$ and edges $\mathcal{E}$ to a FPGA        \\ \cmidrule(l){2-3} 
                            & Feature\_Storing( )    &  Transfer selected vertex feature $\bm{X}_i$  to the local memory of a FPGA           \\ \midrule
\multirow{6}{*}{GNN APIs}   & GNN\_Parameters( )        &    Number of layer $L$, and feature length $f^l$        \\ \cmidrule(l){2-3} 
                            & GNN\_Computation( )       &    The layer operators in GNN model. Specify an off-the-shelf GNN model or "customize"         \\ \cmidrule(l){2-3} 
                            & Scatter( )                &    User defined function, required only if customized layer operator is specified          \\ \cmidrule(l){2-3} 
                            & Gather( )                 &    User defined function, required only if customized layer operator is specified         \\ \cmidrule(l){2-3} 
                            & Update( )                 &     User defined function, required only if customized layer operator is specified        \\ \cmidrule(l){2-3} 
                            & GNN\_Model( )             &    Build GNN model using GNN parameters       \\ \midrule
\multirow{6}{*}{Host APIs}  & FPGA\_Metadata( )       &    FPGA memory bandwidth, number of DSPs, LUTs, etc.         \\ \cmidrule(l){2-3} 
                            & Platform\_Metadata( )       &    PCIe memory bandwidth, number of FPGAs, etc.        \\ \cmidrule(l){2-3} 
                            & Generate\_Design( )         &   Generate hardware design and software design          \\ \cmidrule(l){2-3} 
                             & LoadInputGraph( )       &    File path to read input graph       \\ \cmidrule(l){2-3} 
                            & Start\_training( )        &  Run GNN training           \\ \cmidrule(l){2-3} 
                            & Save\_model( )            &     Save trained GNN model        \\ \bottomrule
\end{tabular}
\end{table*}

\begin{figure}[h]
    \centering
    \includegraphics[width=8.5cm]{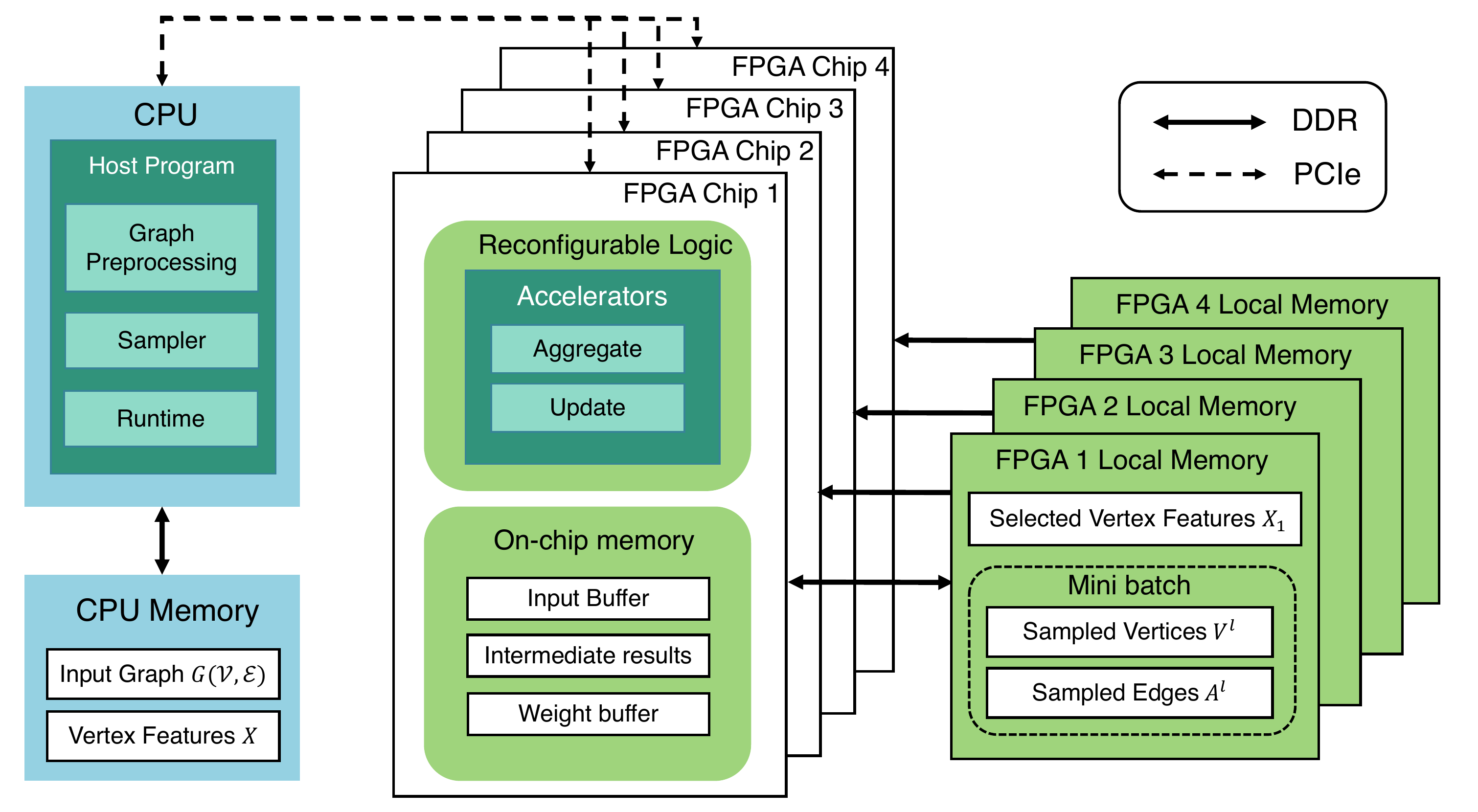}
    \caption{System overview}
     \label{fig:system}
\end{figure} 

HitGNN then parses the input program, and extracts the abstraction of the synchronous GNN training algorithm, which serves as the intermediate representation for the software and accelerator generator to produce the design that runs on the CPU+Multi-FPGA platform. We describe each of the building blocks as follows:

\vspace{0.1cm}
\noindent\textbf{Software generator}: Given the user-specified inputs, the software generator produces a host program. 
During the preprocessing stage, the host program performs graph partitioning, and distributes the vertex features to each FPGA.
During training, the host program performs mini-batch generation and distributes the mini-batches to each FPGA. 
The host program also consists of a runtime system that manages FPGA task scheduling and data communication.

\vspace{0.1cm}
\noindent\textbf{Accelerator generator}: The accelerator generator parses the user input and generates parameterized hardware design using the Kernel Library; the parameters (i.e., accelerator configuration) are determined by the DSE Engine. 
Then, the accelerator generator produces synthesizable hardware for the target FPGA.
\begin{itemize}
    \item \textbf{Kernel Library}: HitGNN provides optimized hardware kernels written in high-level synthesis (HLS) for several widely-used GNN models. User can either implement existing models in the kernel library or customize their own model using the optimized kernel templates.
    \item \textbf{DSE Engine}: The DSE engine takes the platform metadata (e.g., PCIe bandwidth, number of FPGAs) as input, explores the hardware design space, and generates accelerator configurations that optimizes the GNN training throughput (Section \ref{sec:DSE}). 
\end{itemize}
\vspace{-0.3cm}


\subsection{System Overview}\label{sec:system}
Figure \ref{fig:system} depicts the mapping of a synchronous GNN training algorithm onto a CPU+Multi-FPGA platform.
During runtime, the host program on the CPU first performs graph preprocessing, and distributes selected vertex features to each FPGA. 
Then, the host launches GNN training. A sampler produces mini-batches, and the runtime system distributes the mini-batches to the FPGA local memory. 
In the meantime, the FPGAs perform GNN operations. If a vertex feature needed for computation is not present in the FPGA local memory, the FPGA sends a request to host CPU, and the runtime system reads the data from the CPU memory and transfers it to the FPGA (Section \ref{sec:data_comm}). 
After the backpropagation is performed and the gradients are derived, each FPGA sends the gradients back to the host for synchronization. 
The runtime system averages the gathered gradients, and then broadcasts the averaged gradients back to each FPGA to perform a global weight update.

HitGNN performs graph preprocessing and sampling on the host CPU, and performs GNN operations including feature aggregation and feature update on the FPGAs.
Based on this task assignment, the input graph topology $\mathcal{G}$($\mathcal{V}, \mathcal{E}$) and vertex feature $X$ are stored in the host memory for the host CPU to perform graph preprocessing and sampling;
the mini-batch topology $\mathcal{V}^l$ and $\bm{A}^l$ and selected vertex features $X_i$ are stored in the FPGA local memory for the FPGA to perform GNN operations.

\subsection{High-level APIs}\label{sec:api}

Table \ref{tab:API} summarizes the high-level APIs provided to the users to program the synchronous GNN training using Python. 
Listing \ref{lst:program} is an example of mapping a synchronous GNN training algorithm onto the target CPU+Multi-FPGA platform using HitGNN.

\begin{lstlisting}[language=Python, caption=An example user program , label={lst:program}]
### Design Phase ###

'''
Run graph preprocessing programs to produce
V[p], E[p] and X[p]; p = # of FPGAs
'''

for i in range(p): #assign graph data to each FPGA
    Graph_Partition(V[i], E[i], i)
    Feature_Storing(X[i], i)

samp = open('sampling_program')

GNN_comp = GNN_Computation('GCN')
GNN_para = GNN_Parameters(L=2, hidden=[256])
Model = GNN_Model(GNN_comp, GNN_para)

#specify the resoruces of a single super logic region, using Xilinx-U250 as an example
for i in range(p):
    FPGApara[i] = FPGA_Metadata(SLR = 4, DSP=3072, LUT=423000, URAM=320, BRAM= , BW=19.25) 
Platform = Platform_Metadata(BW = bw, FPGA = FPGApara, FPGA_connect = 16)
bitstream, runtime = Generate_Design(Model, samp, Platform) #generate host and accelerator design, return the pointers 

### Runtime Phase ###
Graph = LoadInputGraph('ogbn-product', Path='')
Init(bitstream) # initialize the hardware platform
start_training(runtime, Graph, epochs=10)
save_model()
\end{lstlisting}

\section{CPU+Multi-FPGA Optimizations} \label{sec:opt}
Accelerating synchronous GNN training on a CPU+Multi-FPGA platform suffers from workload imbalance, and high data communication overhead. We describe the optimizations adopted to tackle these challenges in Section \ref{sec:workload} and \ref{sec:data_comm}. 
In addition, we introduce the kernel library in HitGNN, which consists of GNN kernels optimized to achieve high training throughput (Section \ref{sec:library}).

\subsection{Workload Balancing} \label{sec:workload}
In the graph preprocessing stage (Section \ref{sec:method}), the graph is partitioned into $p$ partitions where $p$ is the number of FPGAs on the target platform. 
During the sampling stage, the sampler samples mini-batches from each graph partition, and distributes the mini-batches to each FPGA. 
As introduced in Section \ref{sec:challenge}, the number of vertices and edges within each partition is different;
thus, the number of mini-batches within each graph partition is also different and leads to workload imbalance. 

\begin{figure}[h]
    \centering
    \includegraphics[width=8.5cm]{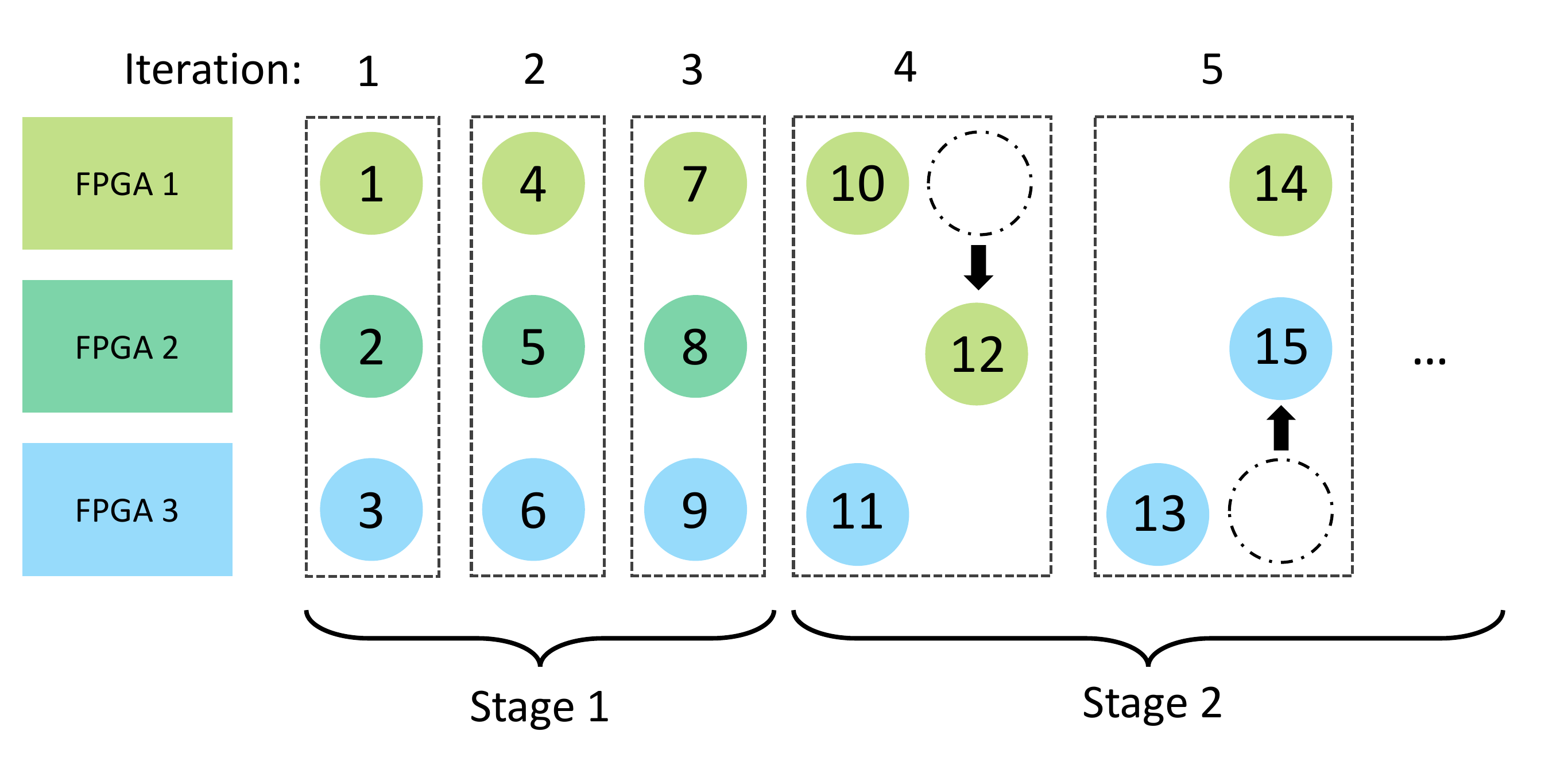}
    \caption{Two-stage task scheduling}
     \label{fig:scheduler}
\end{figure} 

\begin{algorithm}
\caption{Two-Stage Task Scheduling}\label{alg:schedule}
 \textbf{Input}: partitioned graph topology $\mathcal{V}$[i], $\mathcal{E}$[i], where $0\leq$ i $< p$  \\
 \textbf{Output}: mini-batches $b$ and its FPGA assignment
\begin{algorithmic}[1]
\State{\color{blue} Stage 1}
\While{$\#$ of mini-batches in each partition $>$ 0}
\For{i in 0 to \textit{p}}
\State{$b$ = \textbf{Sample}($\mathcal{V}$[i], $\mathcal{E}$[i])}
\State{\textbf{Distribute}($b$, i)} 
{\color{blue}\Comment{Distribute to FPGA $i$}}
\EndFor
\EndWhile
\State{\color{blue} Stage 2}
\State{cnt = 0}
{\color{blue}\Comment{Counter used for round-robin sampling}}
\While{$\#$ of mini-batches in any partition $>$ 0}
\For{i in 0 to \textit{p-1}}
\If{\# of mini-batches in partition $i >$  0}
\State{avail.append(i)}
{\color{blue}\Comment{List of partitions to sample}}
\Else
\State{idle.append(i)} 
{\color{blue}\Comment{List of idle FPGAs}}
\EndIf
\EndFor
\For{i in 0 to avail.length()}
\State{j = avail[cnt \% avail.length()]}
\State{$b$ = \textbf{Sample}($\mathcal{V}$[j], $\mathcal{E}$[j])}
\State{\textbf{Distribute}($b$, avail[i])}
\EndFor

\For{i in 0 to idle.length()} 
\State{j = avail[cnt \% avail.length()]}
\State{$b$ = \textbf{Sample}($\mathcal{V}$[j], $\mathcal{E}$[j])}
{\color{blue}\Comment{Sample extra mini-batch}}
\State{\textbf{Distribute}($b$, idle[i])}
{\color{blue}\Comment{Assign to idle FPGA}}
\State{cnt++} 
\EndFor
\State{avail.clear()}
\State{idle.clear()}
\EndWhile
\end{algorithmic}
\end{algorithm}

We propose a two-stage task scheduler to balance the workload among FPGAs.
We illustrate the idea in Figure \ref{fig:scheduler} with $p=3$. 
The sampler samples each graph partition and produces mini-batches (shown as circles). The color of the mini-batches indicates the graph partition from which it is sampled. The number labeled on each mini-batch indicates the order in which it is produced. The \textit{iteration} labeled in Figure \ref{fig:scheduler} corresponds to the \textit{iteration} in Algorithm \ref{alg:GNN}.

\vspace{0.1cm}
\noindent\textbf{Stage 1}: In stage 1, the sampler is able to sample mini-batches from each graph partition. The task scheduler distributes the mini-batches to each FPGA based on the graph partition the mini-batches are sampled from. 

\begin{figure}[h]
    \centering
    \includegraphics[width=8.5cm]{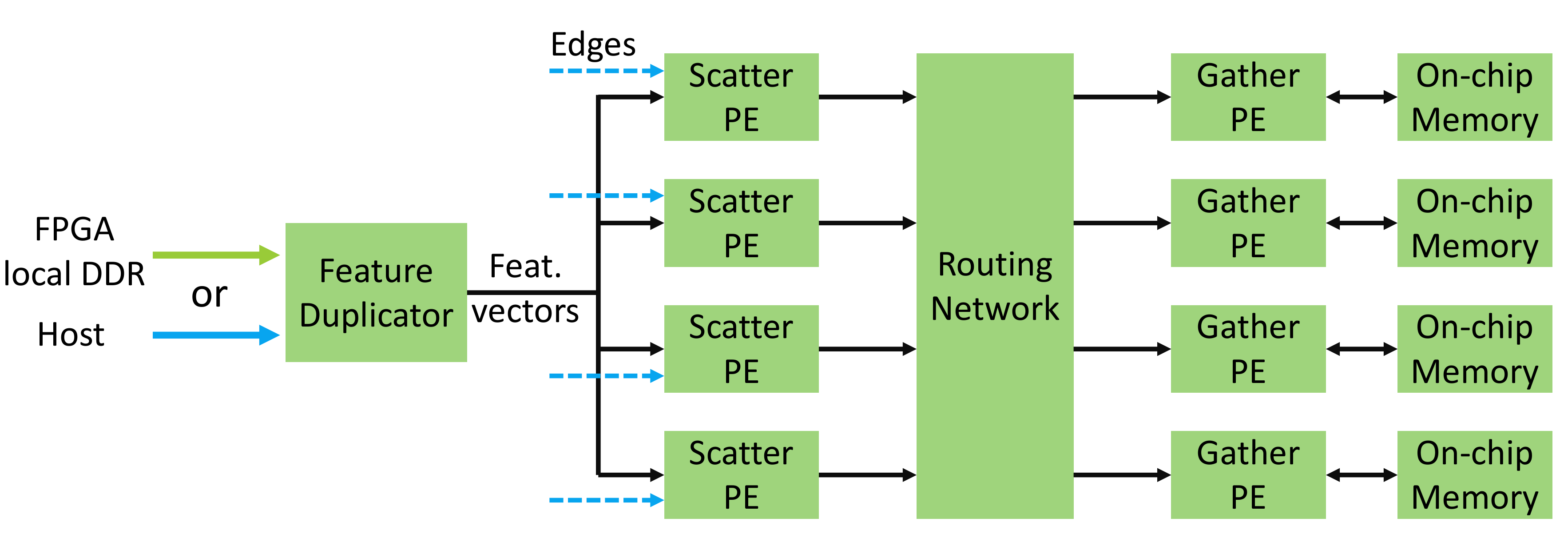}
    \caption{Hardware design of the aggregate kernel}
     \label{fig:scatter}
\end{figure} 

\vspace{0.1cm}
\noindent\textbf{Stage 2}: In stage 2, the mini-batches within some graph partitions have all been executed. 
We show an example in Figure \ref{fig:scheduler} where all the mini-batches within partition 2 have been executed after iteration 3. Thus, for iteration 4, the sampler samples an extra mini-batch from partition 1 to produce 3 mini-batches to perform synchronous SGD. The extra mini-batch (i.e., mini-batch 12) is distributed to FPGA 1 by default, which leads to workload imbalance since FPGA 1 needs to execute 2 mini-batches in iteration 4. 
To balance the workload, the task scheduler distributes the extra mini-batch to an idle FPGA; in this example, the FPGA 2. 
The sampler samples the partitions in a round-robin fashion, so for iteration 5, the sampler moves on and samples a mini-batch from partition 3 and produces mini-batch 13.

We show the general two-stage task scheduling with any number of FPGAs $p$ in Algorithm \ref{alg:schedule}.
With the workload balancing optimization applied, the computations performed remain the same as the original algorithm. 
Using Figure \ref{fig:scheduler} as an example, mini-batch 10, 11, and 12 are executed in iteration 4, regardless of the optimization.


\subsection{Data Communication} \label{sec:data_comm}
In order to train GNN on multiple FPGAs in parallel, the input graph is partitioned and distributed to each FPGA. 
Because of the complex data dependency of graph-structured data, when an FPGA is performing GNN training, the data required may not reside in the local partition.
In this case, the FPGA needs to fetch data from another FPGA which incurs FPGA-to-FPGA communication.
With more FPGAs equipped, the graph is spilt into more partitions, and the data required is more likely to reside in a remote partition.
Previous work have shown that the inter-device communication overhead can easily dominate the synchronous GNN training time when training on a multi-GPU platform \cite{dgcl}.

We show the overview of a CPU+Multi-FPGA platform in Figure \ref{fig:system};
the FPGA-to-FPGA communication is done via a shared memory space in the CPU memory.
In particular, data is first copied from the FPGA memory to the shared memory space and then transferred to another FPGA.
Compared with CPU-FPGA communication, FPGA-to-FPGA communication is much slower because it requires additional data copying to the CPU memory \cite{fpga_comm}.
Thus, we propose to fetch data directly from the CPU memory. 
In particular, whenever a vertex feature required for training is not presented in the FPGA local memory, the FPGA sends a request to the host CPU to fetch the data. 
Fetching vertex features from the CPU memory is feasible because the CPU memory holds the entire graph (Section \ref{sec:system}).
This optimization avoids reading vertex features from another FPGA and thus reduces FPGA-to-FPGA communications. 

Though modern FPGAs support direct communication to another FPGA via Ethernet, this feature is not yet supported on most cloud-FPGA platforms such as Amazon Web Services (AWS) \cite{aws} and Microsoft Azure \cite{azure}. 
In addition,
the controller logic to read data from another FPGA becomes complicated as the platform is equipped with more FPGAs since the controller needs to consider the network topology of FPGA-FPGA connections, and decides the routing to fetch the required data.



\subsection{Optimized Kernel Library}\label{sec:library}

GNN computation is time-consuming because it incurs substantial random memory access. 
In our previous work \cite{hp-gnn}, we develop hardware templates with optimized data-layout and memory organization that can effectively reduce the communication overhead, and therefore achieve high GNN training throughput. 
While these optimizations exploit data parallelism, and increase data reuse, they do not alter the GNN training algorithm.
The hardware template designs follow the computation paradigm mentioned in Algorithm \ref{alg:aggregate-update-paradigm-define}; thus, it is able to support various GNN models. The aggregate kernel adopts a scatter-gather design; The aggregate kernel in Figure \ref{fig:scatter} shows an example of 4 scatter-gather processing elements (PEs), and each PE supports SIMD vector parallelism to process multiple data in one cycle. The update kernel adopts a systolic-array-based design to perform multi-layer perceptron (MLP).

To further reduce development effort, we propose an optimized kernel library, which consists of kernels of widely used GNN models such as GraphSAGE \cite{graphsage}, GCN \cite{gcn}, etc. The kernel library allows users to deploy widely-used GNN models off-the-shelf.  
To support different FPGA platforms, the accelerator configurations in the GNN kernels are parameterized. 
The parameters are automatically determined by the hardware DSE engine (Section \ref{sec:DSE})  after the platform metadata is provided by the user.

\section{Hardware Design Space Exploration} \label{sec:DSE}

HitGNN features a DSE engine that can explore the hardware design space of a CPU+Multi-FPGA platform, and decides the accelerator configurations automatically. In particular, the engine takes the configuration of a mini-batch ($\{|\mathcal{V}^{l}|:0 \leqslant l \leqslant L\}$, $\{|\mathcal{A}^{l}|:1 \leqslant l \leqslant L\}$), GNN hidden dimensions $\{f^{l}:0 \leqslant l \leqslant L\}$, and platform metadata as input, and output a set of parameters that optimizes the GNN training throughput.

\subsection{Resource Utilization Model}\label{sec:resource}
The resource utilization model formulates the hardware resource consumption given a set of accelerator configurations.
In our kernel design, DSPs and LUTs are used the most among the various hardware resources. 
Thus, we model the usage of LUTs and DSPs as our constraints:
\begin{equation}
    \lambda_1\times m  + \lambda_2 \times n  \leq {N}_{DSP}
    \label{eq:DSP}
\end{equation}
\begin{equation}
    \rho_1\times m  + \rho_2 \times n  + \rho_3 \times n \log(n) \leq {N}_{LUT}
    \label{eq:LUT}
\end{equation}
${N}_{\text{DSP}}$ and ${N}_{\text{LUT}}$ denote the available DSPs and LUTs on a single FPGA platform.
For a multi-FPGA platform, each FPGA is constrained by Equation \ref{eq:DSP} and \ref{eq:LUT} independently. All constraints of each FPGA on the target platform needs to be satisfied to produce a valid design. 
We use $m$ to denote the number of processing elements (PEs) in the update kernel, and $n$ to denote the number of scatter-gather PEs in the aggregate kernel (Section \ref{sec:library}).
The coefficients $\lambda_{i}~(1 \leqslant i \leqslant 2)$ and $\rho_{i}~(1 \leqslant i \leqslant 3)$ are constants that indicate the resource consumption for each PE. 
The utilization of DSPs grows linearly as we instantiate more PEs;
as for LUTs, an additional $n\log(n)$ term is introduced to model the LUT overhead of the routing network in the aggregate kernel (Figure \ref{fig:scatter}).

\subsection{Performance Model}\label{sec:performance}
We define the throughput of GNN training as Number of Vertices Traversed Per Second (NVTPS): 
 \begin{equation}
   \text{Throughput} = \frac{\sum_{i=0}^{p}\sum_{l=0}^L|\mathcal{V}^l|}{t_{\text{parallel}}}
    \label{eq:throughput}
\end{equation}
The numerator indicates the total amount of vertices traversed in one iteration. For a multi-FPGA platform with $p$ FPGAs, $p$ mini-batches are computed concurrently. The denominator $t_{\text{parallel}}$ is the parallel execution time of one training iteration, which includes the time for $p$ FPGAs to perform forward propagation, loss calculation, etc. (details are in Algorithm \ref{alg:GNN}) and the time for gradient synchronization. $t_{\text{parallel}}$ can be model as:
\begin{equation}
    t_{\text{parallel}} = \max_{i \in p}\left(t_{\text{execution}}^i\right) + t_{\text{gradient\_sync}}
\end{equation}
Since there are $p$ FPGAs processing in parallel, the parallel execution time is limited by the slowest FPGA; an extra overhead $t_{\text{gradient\_sync}}$ is introduced to for synchronization.

We overlap sampling stage and the GNN computations, so the average execution time on a single FPGA $t_{\text{execution}}$ is estimated as:
\begin{equation}
\begin{split}
    t_{\text{execution}} = \max \left(t_{\text{sampling}}, t_{\text{GNN}}\right)\\
    t_{\text{GNN}} = t_{\text{FP}} + t_{\text{LC}} + t_{\text{BP}} 
\end{split}
\end{equation}
where $t_{\text{\text{GNN}}}$ consists of the execution time of forward propagation $t_{\text{FP}}$, loss calculation $t_{\text{LC}}$, and back propagation $t_{\text{BP}}$. 

The total propagation time $t_{\text{FP}}$ and $t_{\text{BP}}$ is the sum of the execution time of each layer; the execution time of each layer is decided by the task that takes longer to complete since aggregation stage and update stage are pipelined. 
The aggregation stage consists of two tasks: (1) vertex feature loading, and (2) computation. Since the two tasks are pipelined, $t_{\text{aggregate}}$ can be modeled as:
 \begin{equation}
    t_{\text{aggregate}}^l = \max(t_{\text{load}}^l,t_{\text{compute}}^l)
    \label{eq:agg}
\end{equation}

 \begin{equation}
 \resizebox{.9\hsize}{!}{$t_{\text{load}}^l = \frac{|\mathcal{V}^{l-1}|\times \beta\times f^l \times S_{\text{feat}}}{BW_{\text{DDR}}} + \frac{|\mathcal{V}^{l-1}|\times (1-\beta)\times f^l \times S_{\text{feat}}}{BW_{\text{PCIe}}}$}
    \label{eq:load}
\end{equation}

\begin{equation}
    t_{\text{compute}}^l = \frac{|\bm{A}^l|\times f^l }{n\times \text{PE$_{SIMD}$}\times \text{Freq.}}
\end{equation}
We model the vertex feature loading time $t_{\text{load}}^l$ as $\text{(data transferred)}/{\text{(effective bandwidth)}}$. $f^l$ is the feature length, and $S_{\text{feat}}$ is the data size of each feature. $\beta$ is the ratio of fetching data from a local graph partition stored in the local DDR memory (first term of Equation \ref{eq:load}). If the data is in the remote partition, the data is fetched from host via PCIe (second term of Equation \ref{eq:load}).

\renewcommand\algorithmicthen{}

\begin{algorithm}
\caption{Hardware DSE Engine}
\label{alg:DSE}

\begin{algorithmic}[1]
\small
\For{each FPGA}
\For{each die}
\State \textbf{Construct\_Search\_Space( )}
{\color{blue}\Comment{Derive $n_{max},m_{max}$}}
\State{max\_val = 0}
\For{$n=1...n_{max}$}
{\color{blue}\Comment{Exhaustive search}}
\For{$m=1...m_{max}$}
\State{{\color{blue}\#Check resource availability using Eq. (\ref{eq:DSP}), (\ref{eq:LUT})}}
\State{Valid $\gets$ Check\_resource\_availability($n$,$m$)} 
\If{Valid \textbf{and} \textbf{Throughput(}$n$,$m$\textbf{)} $ > $ $max\_val:$}
\State $max\_val$ $\gets$ \textbf{Throughput(}$n$,$m$\textbf{)}
\State \textbf{Save\_configuration(}$n$,$m$\textbf{)}
\EndIf
\EndFor
\EndFor
\EndFor
\EndFor
\end{algorithmic}
\end{algorithm}

We model the compute time as (\# of operations)/(\# of PEs $\times$ kernel frequency). $|\bm{A}^{l}|$ is the number of edges in each layer.
$n$ denotes there are $n$ scatter-gather PEs instantiated in the aggregation kernel. 
Each PE features vector parallelism (Section \ref{sec:library}), and can compute 512-bit of data each cycle; for single-precision floating point data, PE$_{SIMD}$ = 512/32 = 16.

The feature update is modelled as:
\begin{equation}
    t_{\text{update}} = \frac{|V^l|\times f^l \times f^{l+1}}{m\times freq} 
    \label{eq:ns_update}
\end{equation}
We model the $t_{\text{update}}$ as (\# of operations)/(\# of PEs $\times$ kernel frequency). The numerator is the complexity of the multi-layer perceptron and $m$ denotes the number of PEs instantiated in the update kernel.

\subsection{Hardware DSE Engine}
Exploring the design space of a CPU+Multi-FPGA platform is challenging since the design space can be large. 
For example, we can assign some FPGAs to perform feature aggregation, and the others to perform feature update; or, we can assign each FPGA to perform both tasks, but with less parallelism instantiated in both kernels. Our DSE engine adopts the latter approach for two reasons: (1) the bottleneck of GNN computations is vertex feature loading during the feature aggregation, so our design should utilize as much memory bandwidth as possible. That is, all the FPGAs (as opposed to several FPGAs) should utilize their memory bandwidth to perform feature aggregation; and (2) if the FPGAs are able to perform both feature aggregation and feature update independently, the intermediate results can be reused directly; this reduces high volume of data communication among FPGAs. 

The DSE engine explores the design space on each FPGA, and decides the accelerator configurations that optimize the GNN training throughput.
Furthermore, many modern FPGAs consists of multiple dies, and the available resources may vary across dies. Thus, the engine perform DSE for each die to explore the optimal configuration. We assume that each die is connected to one DDR channel (e.g. Xilinx Alveo U250) for simplicity.
The DSE engine first constructs a search space by obtaining the maximum value of $n$ and $m$ separately using Equations (\ref{eq:DSP}) and (\ref{eq:LUT}). Then, the DSE engine performs a parameter sweep through all the possible configurations. 
For each set of configurations, the DSE engine evaluates its throughput using Equation \ref{eq:throughput}, and eventually obtains the optimal design.
We show the steps performed by the DSE engine in Algorithm \ref{alg:DSE}.

\section{{Experiments}}
\subsection{Experimental Setup}\label{sec:setup}
\noindent \textbf{Environments}:
We use our framework to generate GNN training implementations on a CPU+Multi-FPGA platform, and compare the training throughput with a multi-GPU platform. Both the multi-GPU and the CPU+Multi-FPGA platform are built on a dual-socket server. The multi-GPU platform is equipped with 4 GPUs, and the CPU+Multi-FPGA platform is equipped with 4 FPGAs. The GPUs or FPGAs are connected to the host CPU via PCIe.
We list the data of the host CPU, GPUs, and FPGAs in Table \ref{tab:power}.
Note that both the peak performance and memory bandwidth of the FPGA platform are much lower than the GPU platform; thus, the GNN training performance on the CPU+Multi-FPGA platform highly relies on the proposed optimizations. 
The multi-GPU baseline is implemented using Python v3.6, PyTorch v1.11, CUDA v11.3, and PyTorch-Geometric v2.0.3. 
The implementations on the CPU+Multi-FPGA platform are described in Section \ref{sec:implement}.

\begin{table}[!ht]
\centering
\caption{Specifications of the platforms }
\begin{threeparttable}

\begin{adjustbox}{max width=0.47\textwidth}
\renewcommand{\arraystretch}{1.3}
\begin{tabular}{c|ccc}
 \toprule
\textbf{Platforms} & \begin{tabular}[|c|]{@{}c@{}} CPU \\  AMD EPYC 7763 \end{tabular}  & \begin{tabular}[|c|]{@{}c@{}} GPU \\  Nvidia RTX A5000 \end{tabular} & \begin{tabular}[|c|]{@{}c@{}} FPGA \\  Xilinx Alveo U250 \end{tabular}  \\ 
\midrule \midrule
 {Technology}  & TSMC 7 nm+   & Samsung 8 nm & TSMC 16 nm \\ 
{Frequency} & 2.45 GHz  & 2000 MHz & 300 MHz 
      \\ 
{Peak Performance}& 3.6 TFLOPS & 27.8 TFLOPS & 0.6 TFLOPS  \\ 
{On-chip Memory}& 256 MB L3 cache & 6 MB L2 Cache & 54 MB  \\
{Memory Bandwidth}& 205 GB/s & 768 GB/s & 77 GB/s   \\ \bottomrule
\end{tabular}
\end{adjustbox}
\end{threeparttable}
\label{tab:power}
\end{table}
\vspace{-0.2cm}
\begin{small}
\begin{table}[h]
\renewcommand{\arraystretch}{1}
\caption{Statistics of the Datasets and GNN-layer dimensions}
    \centering
    \begin{tabularx}{0.95\columnwidth}{cccYYY}
        \toprule
        \textbf{Dataset} & \textbf{\#Vertices} & \textbf{\#Edges} & $f_{0}$ &  $f_{1}$ &  $f_{2}$\\
        \midrule
        \midrule
        Reddit (RD) & 232,965 & 23,213,838 &  602 & 128 & 41\\
        Yelp (YP) & 716,847 & 13,954,819 &  300 & 128 & 100\\
        Amazon (AM)  & 1,569,960 & 264,339,468 &  200 & 128 & 107\\
        ogbn-products (PR) & 2,449,029 & 61,859,140 &  100 & 128 & 47 \\
        \bottomrule
    \end{tabularx}
    \label{tab: graph-scale}
\end{table}
\end{small}

\noindent \textbf{Synchronous GNN Training Algorithms}:
We evaluate our framework using three representative synchronous GNN training algorithms: DistDGL \cite{distdgl}, PaGraph \cite{pagraph}, and $P^3$ \cite{p3}. 
Note that we only follow how these algorithms perform graph perprocessing (details in Section \ref{sec:method}), and do not implement the  optimizations  in the original works since some optimizations are platform-dependent and cannot be applied to the CPU+Multi-FPGA platform.
Furthermore, HitGNN already has its own optimizations for the CPU+Multi-FPGA platform.

\vspace{0.1cm}
\noindent \textbf{GNN Models and Datasets}:
We evaluate our framework on two well-known GNN models: GraphSAGE \cite{graphsage} and GCN \cite{gcn}. 
We use a 2-layer model with hidden feature size 128, the size of target vertices $|\mathcal{V}^{t}|$ in each mini-batch is set as 1024, and the neighbor sampling size of each layer are 25 and 10; the GNN parameters chosen for evaluation have been widely-used and have shown promising results in many works \cite{graphsage,gcn,graphsaint,ogb,shaDow}.
We choose four widely-used datasets with over ten million edges for evaluation: Reddit, Yelp, Amazon \cite{graphsaint} and ogbn-products \cite{ogb}. 
We list the details of the datasets and GNN-layer dimensions in Table \ref{tab: graph-scale}.

\subsection{Framework Implementation}\label{sec:implement}
HitGNN consists of several building blocks to generate a design that runs on a CPU+Multi-FPGA heterogeneous platform. 
The program parser, DSE engine, software and hardware generator are implemented using Python v3.6, and the accelerator templates are implemented using Xilinx Vitis HLS v2021.2. The host program template is programmed in C++14 with OpenCL library. User interface with HitGNN using the provided APIs (Section \ref{sec:api}). 
In Listing \ref{lst:input}, we provide an example of implementing the three synchronous GNN training algorithms in our experiments; and in Listing \ref{lst:output}, we show part of the generated host program and synthesizable accelerator design. 
$P^3$ requires an extra all-to-all broadcast step at the end of the first GNN layer; we regard this extra step as a special case, and do not provide any APIs to handle it for design simplicity. In Line 14-19 of Listing \ref{lst:output}, we show that this extra step can be easily implemented using built-in functions of the OpenCL library. 
\vspace{0.3cm}

\begin{lstlisting}[language=Python, caption=Example program of the three synchronous GNN training algorithms , label={lst:input}]
### DistDGL preprocessing example ###

'''
Run multi-constraint METIS in DistDGL to produce
V[p], E[p] and X[p] first
'''
for i in range(p): #assign graph data to each FPGA 
    Graph_Partition(V[i], E[i], i)
    Feature_Storing(X[i], i)
    
### PaGraph preprocessing example ###
'''
Run PaGraph's graph partitioning to produce V[p] and E[p], store vertex features with high out degree into array X
'''
for i in range(p): #assign graph data to each FPGA 
    Graph_Partition(V[i], E[i], i)
    Feature_Storing(X, i) #same X for each FPGA

### P3 preprocessing example ###
for i in range(p):
    Graph_Partition(V, E, i) #entire graph topology
    Feature_Storing(X[i], i) #partitioned along feature dimension

samp = open('GraphSAGE_sampler') 

GNN_comp = GNN_Computation('GraphSAGE') 
GNN_para = GNN_Parameters(L=2, hidden=[128])
Model = GNN_Model(GNN_comp, GNN_para)

\end{lstlisting}

\begin{lstlisting}[language=C, caption=Example of generated code, label={lst:output}]
//part of the Host Program
cl::Device devices = xcl::get_xil_devices();
//get all FPGAs on the platform
...//set up and initialization

for(int iter = 0; iter < Layer; iter++) {
// variable "Layer" is specified by user
    buffer_out =  cl::Buffer(CL_MEM_WRITE_ONLY);
    aggregare_krnls.setArg(0, buffer_out);
    ...//set arguments for kernel
    q.enqueueTask(aggregare_krnls);
    q.enqueueTask(update_krnls);
    if(iter == 0){
    //all-to-all broadcast for P^3, manually added 
        for(i=0;i<p;i++) 
            q.enqueueMigrateMemObjects(part_res[i]);
        concat(part_res, layer2ip);
        for(i=0;i<p;i++) 
            q.enqueueMigrateMemObjects(layer2ip[i]);
        }
    }
    q.finish();
    q.enqueueMigrateMemObjects(buffer_out);
    //copy result back to host}
//part of the generated Accelerator Design
read_from_stream(input, tmp_update);
if(tmpupdate.valid == 1){
	index_type dst = tmp_update.dst - dst_offset;
	reg_update = result_buffer[dst];
	
	for (int j = 0; j < 16; j++){
	#pragma HLS unroll = 8 //decide by DSE
		regupdate.data[j] = regupdate.data[j] + tmpupdate.value.data[j];
		}//user-specified aggregate function
		resultbuffer[dst] = regupdate}
\end{lstlisting}

\begin{figure}[h]
    \centering
    \includegraphics[width=6.5cm]{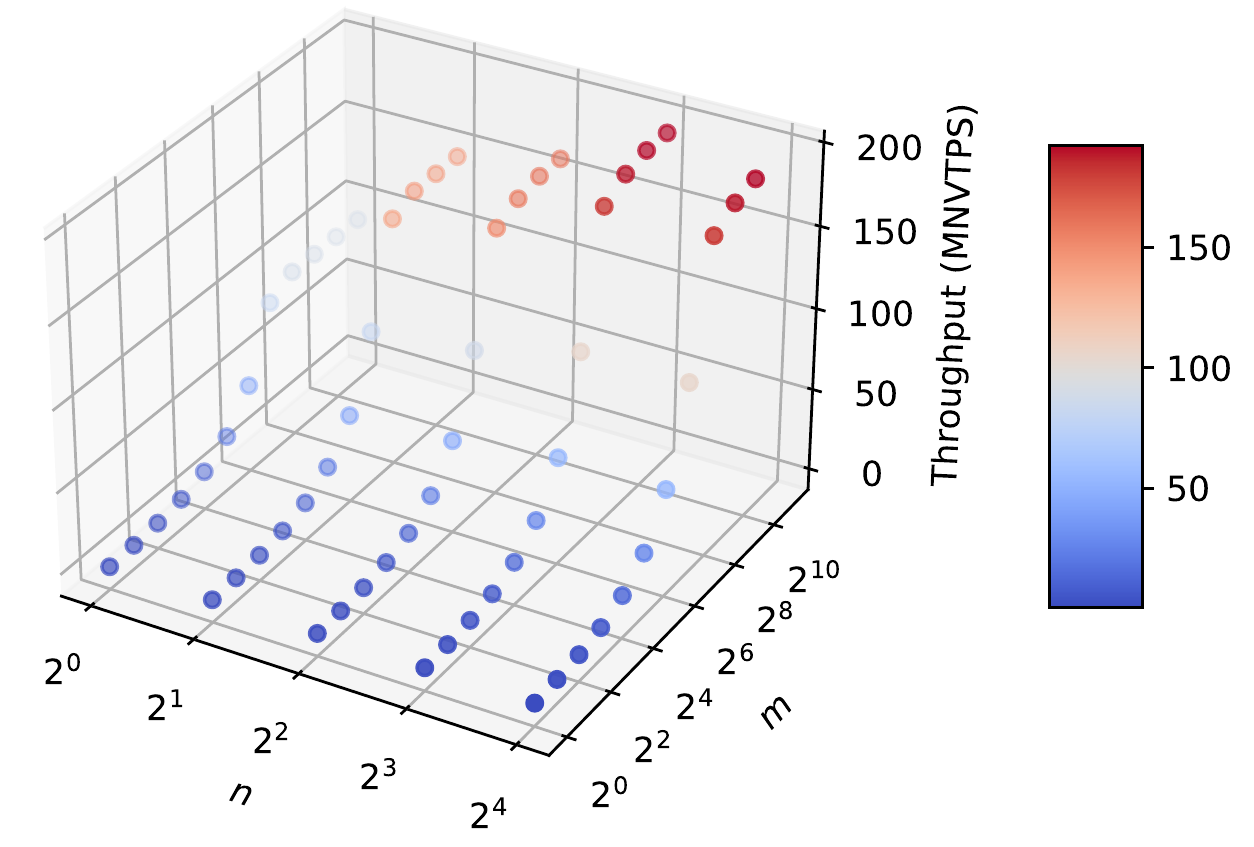}
    \caption{{Result of design space exploration for GraphSAGE}}
     \label{fig:DSE}
\end{figure} 

\subsection{DSE Engine Evaluation}
Given a GNN model, sampling algorithm, and CPU+Multi-FPGA platform metadata, the DSE engine automatically decides the accelerator configurations.
We use \textit{n} to indicate the number of scatter-gather PEs in the aggregate kernel, and use \textit{m} to indicate the number of PEs in the update kernel (Section \ref{sec:DSE}). 
We perform a parameter sweep to explore the FPGA design space as in Algorithm \ref{alg:DSE}. For each design point (\textit{n,m}), the DSE engine estimates the GNN training throughput on the four datasets (Table \ref{tab: graph-scale}), and reports the average throughput. 
We show an example of the DSE result of a GraphSAGE model in Figure \ref{fig:DSE}. DSE on the GCN model also shows similar result.

Since the DSE engine systematically searches through the hardware design space, it is able to find an optimal accelerator configuration.
As shown in Table \ref{tab: resource}, both accelerator configurations (8,2048) and (16,1024) saturates the available hardware resources. 
Since feature aggregation is usually the bottleneck of GNN training, one might choose (16,1024) over (8,2048) to maximize the parallelism of the aggregate kernel, intuitively.
However, as the DSE engine suggests, the configuration (8,2048) leads to higher training throughput instead; this is because our optimized kernel effectively reduces the communication overhead of feature aggregation and shifts the bottleneck to the feature update phase. 
Thus, the configuration (8,2048), which invests more hardware resources in the update kernel delivers higher throughput.

\begin{table}[h]
\caption{Resource utilization and Parallelism}
    \centering
    
    \begin{tabular}{c|cc}
        \toprule
        {Parallelism (\textit{n,m})} & (8,2048) & (16,1024)   \\
        \midrule \midrule
        LUTs  & 72\%  & 65\% \\
        DSPs & 90\% & 56\% \\
        URAM &  48\% & 34\%  \\
        BRAM &  40\%   & 28\% \\\midrule
        Estimated Throughput (NVTPS) & 97.0 M & 92.6 M   \\
        \bottomrule
    \end{tabular}
    
    \label{tab: resource}
\end{table}

\subsection{Performance Metrics}\label{sec:metric}
\begin{itemize}
    \item \textbf{Epoch time}: the time it takes to train one epoch (seconds).
    \item \textbf{Throughput}: we define the training throughput as the Number of Vertices Traversed Per Second (NVTPS).
    \item \textbf{Bandwidth efficiency}: throughput divided by available memory bandwidth of the target platform (NVTPS/(GB/s)). GNN training throughput highly relies on the available memory bandwidth of the platform; since the memory bandwidth varies on different platforms, normalizing the throughput with the available bandwidth provides a fair comparison across different platforms.  
\end{itemize}

\subsection{Cross Platform Comparison}
We compare the performance of a set of designs generated by HitGNN that runs on a CPU+Multi-FPGA platform, with state-of-the-art GNN training implementations using PyTorch-Geometric \cite{pyg} that runs on a multi-GPU platform.
{The CPU+Multi-FPGA platform has four FPGAs and the multi-GPU platform has four GPUs (Section \ref{sec:setup}).}
We show the results in Table \ref{tab:perf}, which uses the metrics defined in Section \ref{sec:metric} for comparison.
We use \textit{GPU} to indicate the multi-GPU baseline, and use \textit{Ours} to indicate the designs generated by HitGNN; we use \textit{GSG} to indicate the GraphSAGE \cite{graphsage} model, and \textit{GCN} to indicate the GCN \cite{gcn} model. 
{Compared with the multi-GPU baseline, HitGNN achieves 2.11$\times$, 2.28$\times$, and 2.34$\times$ speedup for the DistDGL, PaGraph, and $P^3$, respectively.}
This is because HitGNN features optimizations that balance the workload among the FPGAs and reduce FPGA-to-FPGA communication overhead.
To evaluate the effectiveness of our optimizations, we conduct an ablation study.
We first evaluate the performance of a baseline design, and then gradually add the workload balancing (WB) optimization and the data communication (DC) optimization to the design.
We show the evaluation in Table \ref{tab:improve} which uses the DistDGL algorithm as an example, the two optimizations can deliver up to 66\% throughput improvement in total. 
Furthermore, our highly-optimized GNN kernels allows HitGNN to achieve 13.4$\times$-14.9 $\times$ bandwidth efficiency than the multi-GPU platform w.r.t. the geometric mean.
Due to the superior bandwidth efficiency, HitGNN achieves up to 4.26$\times$ speedup using only 0.16$\times$ memory bandwidth of the multi-GPU platform.

\renewcommand{\arraystretch}{1.1}
\begin{table*}[]
\caption{Cross platform comparison}
\label{tab:perf}
\begin{adjustbox}{max width=0.9\textwidth,center}
\begin{tabular}{lccrrrrrrrrc}
\toprule
\multicolumn{2}{l}{\multirow{2}{*}{}}                                                                       & Dataset & \multicolumn{2}{c}{Reddit}                        & \multicolumn{2}{c}{Yelp}                          & \multicolumn{2}{c}{Amazon}                        & \multicolumn{2}{c}{ogbn-products}                 & \multicolumn{1}{c}{\multirow{2}{*}{Geo. Mean}} \\ \cline{3-11}
\multicolumn{2}{l}{}                                                                                        & Model   & \multicolumn{1}{c}{GCN} & \multicolumn{1}{c}{GSG} & \multicolumn{1}{c}{GCN} & \multicolumn{1}{c}{GSG} & \multicolumn{1}{c}{GCN} & \multicolumn{1}{c}{GSG} & \multicolumn{1}{c}{GCN} & \multicolumn{1}{c}{GSG} & \multicolumn{1}{c}{}                           \\ \midrule \midrule
                            

\multirow{6}{*}{DistDGL\cite{distdgl}} & \multirow{2}{*}{Epoch time (s)}                                                      & GPU     & 1.29                    & 1.34                    & 1.75                    & 1.79                    & 4.19                    & 4.34                    & 5.03                    & 5.38      & -              \\ \cline{3-12} 
                         &                                                                                  & Ours    & 0.62                    & 0.77                    & 0.63                    & 0.87                    & 1.14                    & 1.72                    & 3.06                    & 4.31       & -             \\ \cline{2-12} 
                         & \multirow{2}{*}{\begin{tabular}[c]{@{}c@{}}Throughput \\ (NVTPS) \end{tabular}}                                                      & GPU     & 15.6 M               & 15.1 M                & 21.6 M                & 21.1 M               & 22.6 M               & 21.8 M                & 97.5 M                & 91.2 M    & 28.8 M (1.00$\times$)            \\ \cline{3-12} 
                         &                                                                                  & Ours    & 32.5 M                & 26.2 M                & 59.9 M                & 43.4 M                & 83.1 M                & 55.1 M                & 160 M                & 114 M     &     60.7 M (2.11$\times$)       \\ \cline{2-12} 
                         & \multirow{2}{*}{\begin{tabular}[c]{@{}c@{}}BW efficiency \\ (NVTPS/(GB/s))\end{tabular}} & GPU     & 4.77 K                & 4.59 K                & 6.58 K                & 6.44 K               & 6.90 K               & 6.66 K                & 29.8 K                & 27.8 K          & 8.78 K (1.00$\times$)      \\ \cline{3-12} 
                         &                                                                                  & Ours    & 63.4 K                & 51.1 K                & 117 K                & 84.6 K                & 162 K                & 107 K                & 313 K                & 222 K     & 118 K (13.4$\times$)        \\ \midrule
\multirow{6}{*}{PaGraph\cite{pagraph}} & \multirow{2}{*}{Epoch time (s)}                                                      & GPU     & 1.21                    & 1.25                    & 1.66                    & 1.69                    & 4.04                    & 4.16                    & 4.61                    & 4.89       & -             \\ \cline{3-12} 
                         &                                                                                  & Ours    & 0.53                    & 0.67                    & 0.54                    & 0.77                    & 0.98                    & 1.53                    & 2.64                    & 3.82     & -               \\ \cline{2-12} 
                         & \multirow{2}{*}{\begin{tabular}[c]{@{}c@{}}Throughput \\ (NVTPS) \end{tabular}}                                                     & GPU     & 16.7 M                & 16.1 M               & 22.7 M                & 22.3 M                & 23.4 M               & 22.8 M                & 106 M                & 100 M      & 30.6 M   (1.00$\times$)        \\ \cline{3-12} 
                         &                                                                                  & Ours    & 38.1 M               & 30.1 M                & 69.9 M                & 49.0 M                & 96.6 M                & 61.9 M                & 186 M                & 128 M     & 69.8 M  (2.28$\times$)          \\ \cline{2-12} 
                         & \multirow{2}{*}{\begin{tabular}[c]{@{}c@{}}BW efficiency \\ (NVTPS/(GB/s))\end{tabular}} & GPU     & 5.09 K                & 4.92 K                & 6.94 K                & 6.82 K                & 7.15 K                & 6.95 K               & 32.5 K                & 30.6 K   & 9.35 K  (1.00$\times$)            \\ \cline{3-12} 
                         &                                                                                  & Ours    & 74.2 K                & 58.7 K                & 136 K                & 95.6 K               & 188 K                & 121 K                & 362 K                & 250 K    &    136 K  (14.6$\times$)        \\ \midrule
\multirow{6}{*}{P$^3$\cite{p3}}      & \multirow{2}{*}{Epoch time (s)}                                                      & GPU     & 1.24                    & 1.32                    & 1.82                    & 1.79                    & 4.43                   & 4.41                    & 5.21                    & 5.43      & -              \\ \cline{3-12} 
                         &                                                                                  & Ours    & 0.53                    & 0.70                     & 0.55                    & 0.82                    & 1.04                    & 1.70                     & 2.71                    & 4.13        & -            \\ \cline{2-12} 
                         & \multirow{2}{*}{\begin{tabular}[c]{@{}c@{}}Throughput \\ (NVTPS) \end{tabular}}                                                     & GPU     & 16.3 M               & 15.3 M                & 20.7 M               & 21.1 M                & 21.4 M                & 21.5 M                & 94.2 M                & 90.4 M      & 28.4 M   (1.00$\times$)        \\ \cline{3-12} 
                         &                                                                                  & Ours    & 38.1 M                & 28.8 M                & 68.6 M                & 46.0 M                & 91.1 M                & 55.7 M                & 181 M                & 119 M      & 66.4 M  (2.34$\times$)         \\ \cline{2-12} 
                         & \multirow{2}{*}{\begin{tabular}[c]{@{}c@{}}BW efficiency \\ (NVTPS/(GB/s))\end{tabular}} & GPU     & 4.96 K                & 4.66 K                & 6.33 K                & 6.44 K                & 6.52 K                & 6.55 K                & 28.7 K               & 27.6 K     & 8.67 K   (1.00$\times$)        \\ \cline{3-12} 
                         &                                                                                  & Ours    & 74.2 K               & 56.2 K                & 134 K                & 89.7 K                & 178 K                & 109 K                & 353 K                & 232 K      & 129 K    (14.9$\times$)       \\ \bottomrule

\end{tabular}
\end{adjustbox}
\end{table*}

\begin{table}[!ht]
\renewcommand{\arraystretch}{1}
\setlength{\tabcolsep}{10pt}
\centering
\caption{Throughput improvement due to optimizations}

\begin{threeparttable}

\begin{tabularx}{\columnwidth}{c|XXXc}
 \toprule
{Data-Model} & Baseline & WB & WB+DC & Speedup   \\ 
\midrule
\midrule
{RD-GCN} &19.9 M  & 22.7 M & 32.5 M  & 63\%  \\ 
{RD-GSG} &16.9 M  & 19.2 M & 26.2 M  & 55\%\\
{YP-GCN} &36.4 M  & 41.9 M & 59.9 M  & 65\%    \\ 
{YP-GSG} &28.5 M  & 32.8 M & 43.4 M  & 52\%\\
{AM-GCN} &50.8 M  & 59.6 M & 84.1 M  & 64\%\\ 
{AM-GSG} &36.5 M  & 42.9 M & 55.1 M  & 51\%\\
{PR-GCN} &96.7 M  & 113  M & 160 M   & 66\%  \\ 
{PR-GSG} &73.8 M  & 86.5 M & 114 M   & 54\% \\ \bottomrule

\end{tabularx}
\end{threeparttable}
\label{tab:improve}
\end{table}

\vspace{-0.2cm}
\subsection{Scalability}

\begin{figure}[t]
    \centering
    \includegraphics[width=7.2cm]{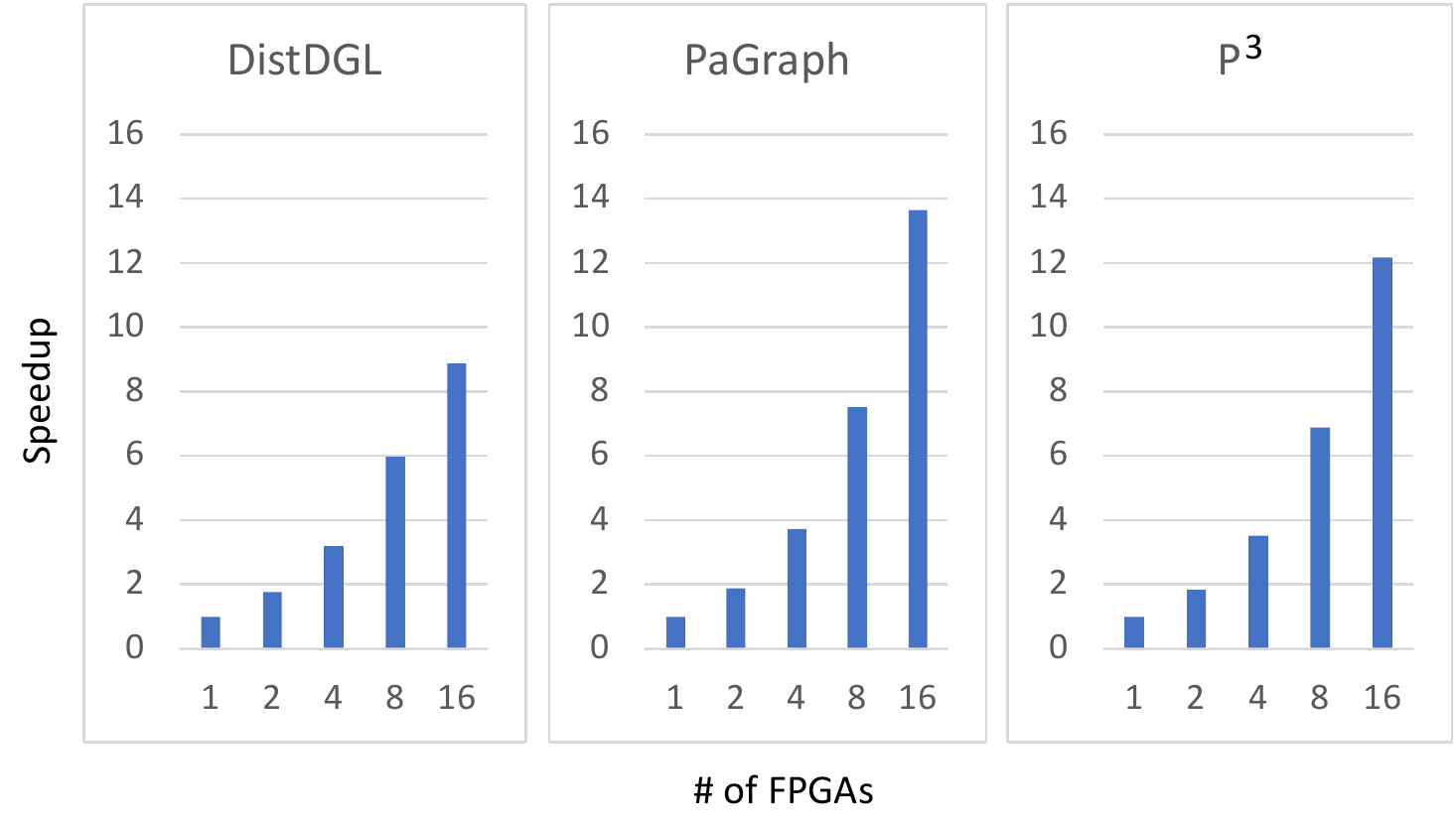}
    \caption{Scalability evaluation}
     \label{fig:scale}
\end{figure}

{We build a CPU+Multi-FPGA platform simulator to evaluate the scalability of HitGNN. 
The simulator estimates the training performance given the synchronous GNN training algorithm, GNN model, and platform metadata.}
To verify the simulator, we first implement the host program and the hardware kernels. Then, we measure the host program execution time, and the post-synthesis kernel execution time to fine-tune the simulator. 
We show the speedup compared with a single FPGA for each algorithm in Figure \ref{fig:scale}.
By reducing the FPGA-to-FPGA communication overhead, and balancing the workload, HitGNN achieves scalable speedup.
The scalability of HitGNN is limited by the CPU memory bandwidth. 
This is because FPGAs fetch data from the CPU memory if the data required is not presented in the FPGA local memory (Section \ref{sec:data_comm}). 
Using the EPYC 7763 CPU as an example, the CPU memory (bandwidth = 205 GB/sec.) can serve up to 205/16 = 12.8 FPGAs without saturating the CPU memory bandwidth, where the denominator is the bandwidth of a single CPU-FPGA connection via PCIe;
if more FPGAs are added to the platform, the scalability of the speedup starts to decrease since the CPU memory bandwidth is gradually saturated. 
In Figure \ref{fig:scale}, we show that the throughput of HitGNN scales almost linearly up to 16 FPGAs.

\section{Conclusion}
In this paper, we proposed HitGNN, a general framework to generate high-throughput GNN training implementation on a given CPU+Multi-FPGA heterogeneous platform. 
HitGNN features various optimizations to accelerate synchronous GNN training on the CPU+Multi-FPGA platform. 
The optimizations of HitGNN are general enough to be applied to various synchronous GNN training algorithms; in addition, these optimizations do not alter the training algorithm, thus, they do not affect the model accuracy or the convergence rate.
The implementations generated by HitGNN achieved up to $27\times$ bandwidth efficiency compared with state-of-the-art multi-GPU baseline, and thus achieved up to $4.26\times$ throughput using much less compute power and memory bandwidth than the multi-GPU baseline.
We show that HitGNN can achieve scalable speedup to 16 FPGAs, with the limiting factor being the CPU memory bandwidth.
In the future, we plan to further improve the scalability of HitGNN by exploiting
data prefetching techniques and high-speed FPGA interconnection network (e.g., SmartNIC) to relieve the stress on the CPU memory bandwidth.

\ifCLASSOPTIONcompsoc
  \section*{Acknowledgments}
\else
  \section*{Acknowledgment}
\fi

{This work has been supported by the U.S. National Science Foundation (NSF) under grants SaTC-2104264 and OAC-2209563, and the DEVCOM Army Research Lab (ARL) under grant W911NF2220159.}

\ifCLASSOPTIONcaptionsoff
  \newpage
\fi

\vspace{-0.5cm}
\begin{IEEEbiography}[{\includegraphics[width=1in,height=1.25in,clip,keepaspectratio]{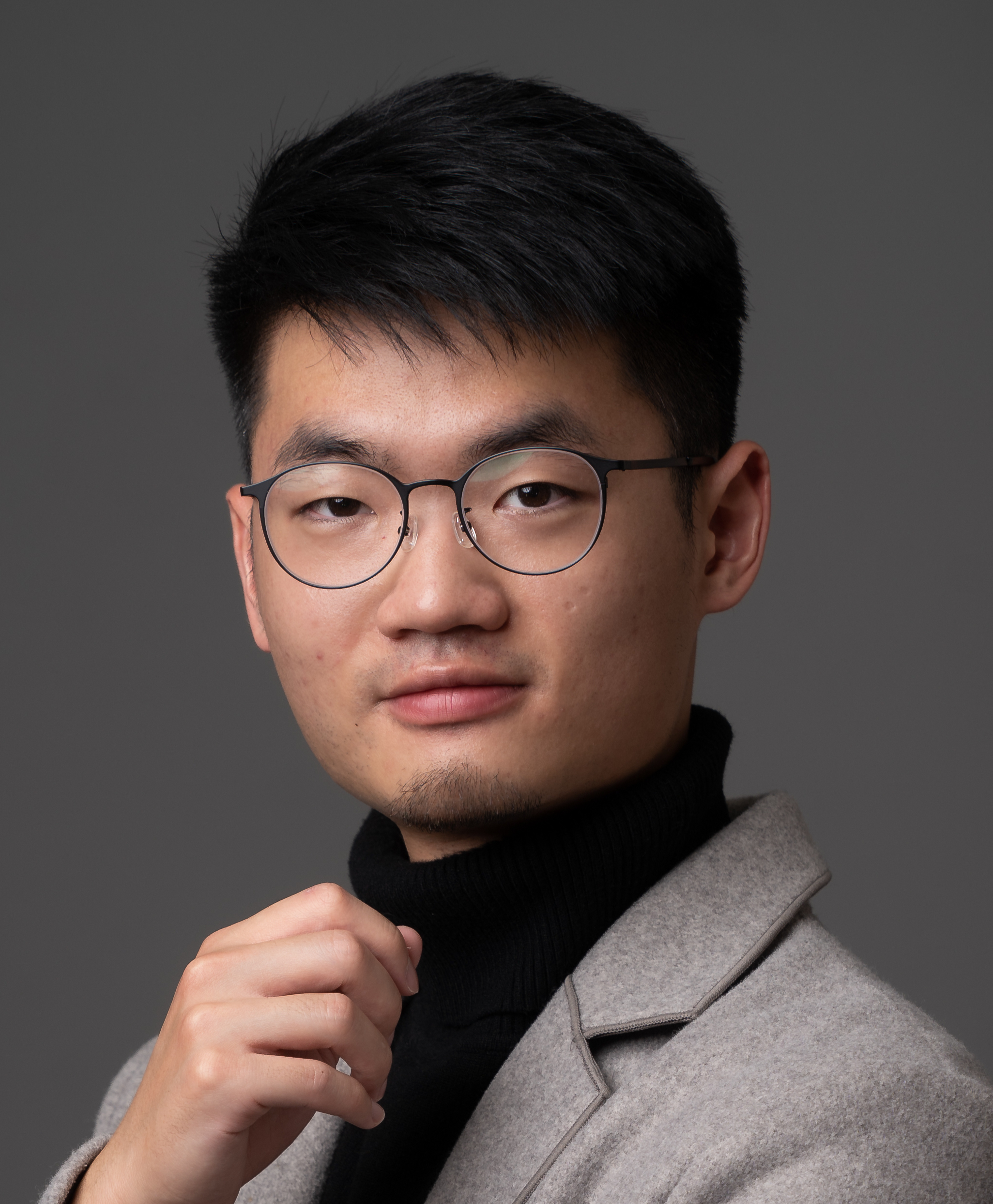}}]{Yi-Chien Lin} received his BS degree in electrical engineering from the National Taiwan University, in 2020, and is currently a Ph.D. student in Electrical Engineering at the University of Southern California. He is a recipient of the Viterbi School of Engineering Graduate School Fellowship at the Ming Hsieh Department of Electrical and Computer Engineering. 
His research interests include heterogeneous computing, hardware acceleration, and graph machine learning.

\end{IEEEbiography}
\vskip -3\baselineskip plus -1fil
\begin{IEEEbiography}[{\includegraphics[width=1in,height=1.25in,clip,keepaspectratio]{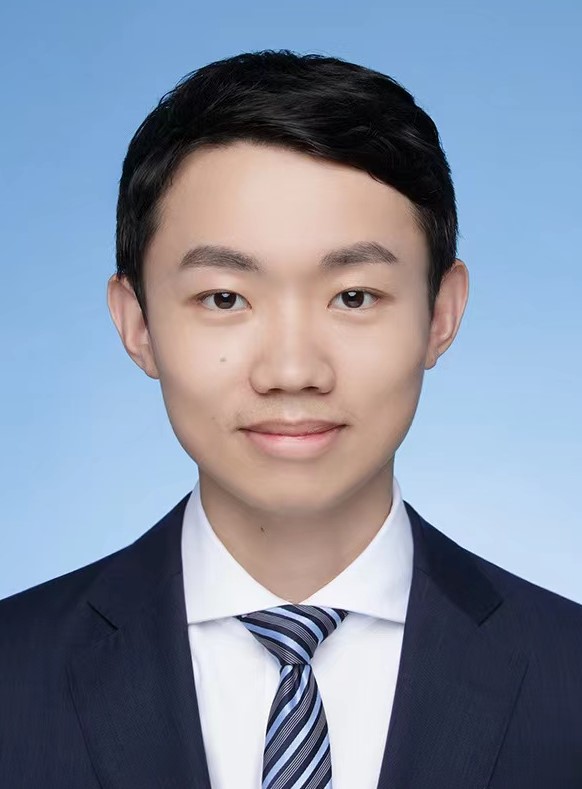}}]{Bingyi Zhang}
received the BS degree in microelectronics from Fudan University, in 2017, and the MS degree in Integrated Circuit Engineering from Fudan University. He is pursuing the PhD degree in computer engineering at the University of Southern California (USC). His research interests include parallel computing, digital signal processing, digital circuit design. He focuses on accelerating graph-based machine learning on FPGA platform. 
\end{IEEEbiography}
\vskip -3\baselineskip plus -1fil
\begin{IEEEbiography}[{\includegraphics[width=1in,height=1.25in,clip,keepaspectratio]{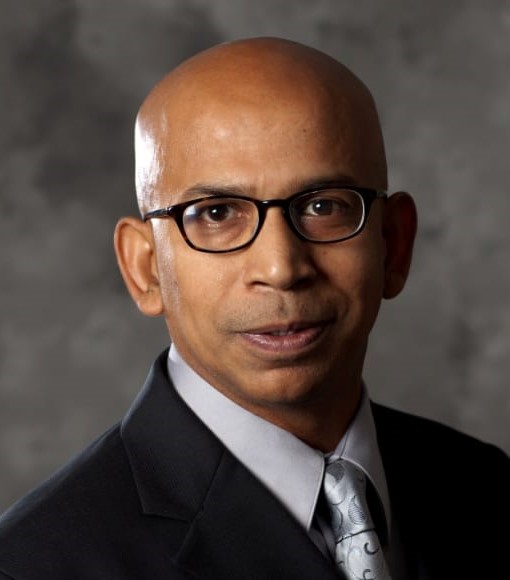}}]{Viktor K. Prasanna}
 received the BS degree in electronics engineering from Bangalore University, the MS
degree from the School of Automation, Indian Institute
of Science, and the Ph.D. degree in computer science from Pennsylvania State University. He is Charles
Lee Powell chair in engineering in the Ming Hsieh
Department of Electrical and Computer Engineering and professor
of computer science with the University of Southern California (USC). His research interests include
high performance computing, parallel and distributed
systems, reconfigurable computing, and embedded systems. He is the executive director of the USC-Infosys Center for Advanced
Software Technologies (CAST) and was an associate director of the USC Chevron
Center of Excellence for Research and Academic Training on Interactive Smart
Oilfield Technologies (Cisoft). He also serves as the director of the Center
for Energy Informatics, USC. He served as the editor-in-chief of the IEEE
Transactions on Computers during 2003–06. Currently, he is the editor-in-chief
of the Journal of Parallel and Distributed Computing. He was the founding
chair of the IEEE Computer Society Technical Committee on Parallel Processing.
He is the steering chair of the IEEE International Parallel and Distributed
Processing Symposium (IPDPS) and is the steering chair of the IEEE International
Conference on High Performance Computing (HiPC). He received the 2009
Outstanding Engineering Alumnus Award from the Pennsylvania State University. He received the W. Wallace McDowell Award from the IEEE Computer
Society, in 2015 for his contributions to reconfigurable computing. His work
on regular expression matching received one of the most significant papers
in FCCM during its first 20 years award in 2013. He is a fellow of the
IEEE, the ACM, and the American Association for Advancement of Science
(AAAS).
He is a member of Academia Europaea.
\end{IEEEbiography}



\end{document}